\documentclass[preprint]{aastex}
\usepackage{amsmath}
\usepackage{graphicx}
\shorttitle{Ambipolar Diffusion in Action}
\shortauthors{Chen \& Ostriker}
\begin{document}
\title{Ambipolar Diffusion in Action: Transient C shock Structure and Prestellar Core Formation}
\author{Che-Yu Chen and Eve C. Ostriker}
\affil{Department of Astronomy, University of Maryland, College Park, MD 20742}
\email{cychen@astro.umd.edu; ostriker@astro.umd.edu}
\begin{abstract}
We analyze the properties of steady and time-dependent C shocks under conditions prevailing in giant molecular clouds. For steady C shocks, we show that ionization equilibrium holds and use numerical integrations to obtain a fitting formula for the shock thickness mediated by ambipolar diffusion, $L_\mathrm{shock}\propto {n_0}^{-3/4}{v_0}^{1/2}{B_0}^{1/2}{\chi_{i0}}^{-1}.$ Our formula also agrees with an analytic estimate based on ion-neutral momentum exchange. Using time-dependent numerical simulations, we show that C shocks have a transient stage when the neutrals are compressed much more strongly than the magnetic field. The transient stage has a duration set by the neutral-ion collision time, $t_\mathrm{AD} \sim L_\mathrm{shock}/v_\mathrm{drift}\sim 0.1-1$~Myr. This transient creates a strong enhancement in the mass-to-magnetic flux ratio. Under favorable conditions, supercritical prestellar cores may form and collapse promptly as a result of magnetic flux loss during the transient stage of C shocks.
\end{abstract}

\section{Introduction}
\label{sec:intro}

Within giant molecular clouds (GMCs), dense gravitationally bound cores form and collapse to create protostars \citep{1987ARA&A..25...23S, 2007ARA&A..45..565M, 2009sfa..book..254A}. Supersonic turbulence is believed to strongly affect the core formation and evolution processes, with post-shock dense regions the most susceptible to collapse (see \cite{2011ApJ...729..120G} and references therein). These processes can be modified significantly by the interstellar magnetic field. Sufficiently strong magnetic fields, if they are well-coupled to the gas, can entirely prevent collapse \citep{1956MNRAS.116..503M}; this can be expressed in terms of a maximum ratio of mass to magnetic flux, $M/\Phi$, or of surface density to magnetic field strength, $\Sigma/B$ \citep{1976ApJ...210..326M, 1978PASJ...30..671N}. However, in a partially-ionized medium, magnetic fields are coupled to the neutrals only through ion-neutral collisions. This ambipolar drift modifies the dynamical effect of magnetic fields on the neutral gas \citep{1979ApJ...228..475M}, in particular altering the character of shocks \citep{1993ARA&A..31..373D}.

In ideal MHD, the fluid and magnetic fields are perfectly coupled by assumption. When flow velocities exceed the relevant signal propagation speeds for a magnetized medium, discontinuities representing shock fronts (jump shock or J-type shock) can form. The compression ratio is parametrized by the particle density, inflow velocity, and magnetic field \citep[e.g.,][]{1992phas.book.....S}. However, in lightly ionized clouds, velocity differences that would produce a J shock in ideal MHD are small compared to the magnetic signal speed (``Alfv\'{e}n speed") in the ionized medium, $v_{\mathrm{A},i} = B/\sqrt{4\pi\rho_i}$. Ions and magnetic fields therefore smoothy transition between upstream and downstream conditions without discontinuities. As a result of the ion-neutral drag forces, the transition in the neutrals is also modified and all physical quantities vary smoothly in the shock region, forming a continuous (C-type) shock \citep{1980ApJ...241.1021D}. In a steady C shock, upstream and downstream values of the neutral density, velocity, and magnetic field are the same as for a J shock. Thus, upstream and downstream values of the mass-to-magnetic flux ratio (per unit length parallel to the shock) are the same. Many studies of C shocks have investigated their formation \cite[e.g.][]{1997A&A...326..801S}, structure \cite[e.g.][]{1995ApJ...442..726M}, and stability \cite[e.g.][]{1990MNRAS.246...98W, 1997ApJ...487..271S}, as well as detailed chemical and emission properties \citep[e.g.][]{1983ApJ...264..485D, 1997IAUS..182..199P}.

Ambipolar diffusion may play a key role in the star-forming process. In the traditional picture, quasi-static prestellar cores form by gravitationally-driven ambipolar diffusion in magnetically-supported clouds (see review by \cite{2009sfa..book..254A}). For a star to form out of gas that is initially strongly magnetized, dense cores must lose magnetic support so that gravitational collapse can take place \citep{1978prpl.conf..209M, 1989ApJ...342..834L}. If the magnetic pressure in a gravitationally-confined core exceeds that in its surroundings, the gradient in magnetic pressure makes the magnetic field (and ions) tend to expand. The neutrals will be left behind as a supercritical core as the magnetic field diffuses outward \citep{1979PASJ...31..697N}. More realistically, \cite{1979ApJ...228..475M} argued that a cloud does not need to lose magnetic flux as a whole to collapse. Rather, ambipolar diffusion redistributes the mass within dense clumps, with the neutrals diffusing inward while the magnetic field threading the outer region is left behind. The duration of the ambipolar diffusion process can be considerably longer (up to a factor of $10$) than the gravitational free-fall timescale $t_\mathrm{ff}$, although the evolution is more rapid if cores are initially closer to critical \citep[e.g.][]{1999osps.conf..305M, 2001ApJ...547..272C}.

Observationally, the prestellar core lifetime can be estimated by calculating the ratio of the number of cores with embedded young stellar objects (YSOs) to the number of prestellar cores, which should be comparable to the ratio of protostar lifetime to the prestellar core lifetime \citep{1999ApJS..123..233L}. Several studies have suggested a prestellar core lifetime of $\sim 10^6$~yr, or $(2-5) t_\mathrm{ff}$ \citep{2007prpl.conf...33W, 2009ApJS..181..321E}. This value is much lower than expected from the magnetic-dominated model. In addition, in the turbulence-controlled regime where the magnetic field and ambipolar diffusion play minor roles \citep{2004RvMP...76..125M}, ideal MHD simulations have shown that cores only live for $(1-2) t_\mathrm{ff}$ \citep[e.g.][]{2005ApJ...618..344V}, after which they either collapse or re-expand. This would not permit an extended period of ambipolar diffusion. 

Several studies have suggested that turbulence in GMCs can accelerate ambipolar diffusion and star formation, by introducing large local gradients and nonlinearities. Considering small-scale fluctuations in a background field, \cite{2002ApJ...570..210F} analytically showed that turbulence can enhance the ambipolar diffusion rate by a factor of $2-3$ for typical conditions in GMCs. \cite{2004ApJ...603..165H} investigated this problem numerically in a 2.5-dimensional geometry and concluded that the enhanced diffusion rate must be balanced against large-scale compressive flows. Independently, \cite{2004ApJ...609L..83L} and \cite{2005ApJ...631..411N} noted that the failure of the standard theory to predict core formation timescales indicates that dense clumps may not have formed quasi-statically through ambipolar diffusion. By performing two-dimensional simulations of magnetized sheetlike clouds, they found that with sufficiently strong turbulence, dense filaments can form from magnetic-field-dominated clouds in one turbulence crossing time ($t\sim 10^6$~yr). 

Turbulence-accelerated, magnetically-regulated star formation was studied by \cite{2008ApJ...679L..97K} using three-dimensional simulations, including self-gravity and adopting hydrostatic equilibrium in the vertical direction as an initial condition. More recently, \cite{2011ApJ...728..123K} conducted a parameter study of fragmentation in magnetically subcritical clouds regulated by ambipolar diffusion and nonlinear turbulent flows. They concluded that the core formation time is strongly affected by the turbulence speed and the density in compressed region. These and other recent simulations with both strong turbulence and ambipolar diffusion \citep[e.g.,][]{2008ApJ...687..354N} are consistent with observations in terms of the core evolution time, the relatively low efficiency of star formation ($\sim 3-6\%$, see \cite{2009ApJS..181..321E}), and the core structure (subsonic infall motions, see \cite{1999ApJ...526..788L}). However, the fundamental physical process driving core formation via turbulence-enhanced ambipolar diffusion, as well as its dependence on environmental parameters, still remain unclear.

To investigate this problem, we consider the simplest possible time-dependent problem with large spatial gradients: a one-dimensional high-speed converging flow that shocks. In order to clearly distinguish the effect of ambipolar diffusion from other dynamics, we neglect the self-gravity of the gas. We also focus on the simplified case in which the inflow velocity is perpendicular to the magnetic field lines; more general geometry (i.e.~oblique shocks) is discussed in Appendix~\ref{sec:appendix}.

When gas is compressed by converging flow, the neutrals are pushed to accumulate downstream. The ion density and magnetic field strength, however, will be only moderately enhanced since the magnetic pressure resists strong compression. These lagging ions exert a drag force on neutrals, reducing the streaming of neutrals into the post-shock region. The momentum exchange between neutrals and ions speeds up ions, increases the compression of the magnetic field, and reduces the post-shock density of the neutrals. Over time, a steady C shock develops. However, at early stages, for an interval comparable to the neutral-ion collision time, the neutrals do not experience drag forces from the ions \citep{2007MNRAS.382..717R, 2009MNRAS.395..319V, 2010A&A...511A..41A}. As a consequence, the initial shock for the neutrals is essentially unmagnetized, and the neutrals can be very strongly compressed. If the gravitational collapse timescale is sufficiently short, and a dense enough layer of gas builds up, the magnetically-supercritical region may be able to collapse gravitationally before a steady C shock structure forms. The transient ambipolar diffusion process in shocks may help to explain the physics of turbulence-accelerated, magnetically-regulated star formation. 

In this paper, we first revisit the steady-state structure of C-type shocks in conditions appropriate for GMCs, in particular allowing for varying ionization fraction. By fitting the results of steady one-dimensional solutions, we obtain an expression for the C shock thickness as a function of the upstream density, the velocity, the magnetic field, and the ionization fraction. These C shock thicknesses are comparable to, or exceed, the size of observed cores. We then consider time-dependent shocks, which we follow by implementing ambipolar diffusion in the MHD code, \textit{Athena}. Our simulations suggest that under some circumstances, transient C shocks make it possible for a magnetically subcritical cloud to form supercritical dense cores, which would then be able to collapse promptly. We show more generally that the mass-to-flux ratio is significantly increased by ambipolar diffusion in transient post-shock regions, compared to the value that would hold under ideal MHD or in a steady C shock.

This paper is organized as follows. The model and the governing equations are described in Section~\ref{sec: eqns}. In Section~\ref{sec: thickness} we investigate the structure of steady C shocks, and obtain (analytically and numerically) an explicit formula for the dependence of shock thickness on environmental parameters. In Section~\ref{sec: formation} the time-dependent numerical method is described, and we show that in the transient early development of C shocks, the post-shock ratio of density to magnetic field is very large. In Section~\ref{sec: criticality}, we discuss mass-to-flux ratios of shocked gas, which we use to quantify the effect of ambipolar diffusion. A parameter study of the duration and effect of transient C shocks is presented in Section~\ref{sec: core}. We summarize our conclusions in Section~\ref{sec: summary}.

\section{Dynamical Equations and Model Parameters}
\label{sec: eqns}
\subsection{Basic Equations}
For a partially ionized medium with a drag force $\mathbf{f}_d$ between ions and neutrals, the neutral fluid equations are
\begin{subequations}
\begin{align}
\frac{\partial\rho_n}{\partial t} &+ \mathbf{\nabla}\cdot\left(\rho_n\mathbf{v}_n\right) = 0,\label{mCon}\\
\rho_n \bigg[\frac{\partial\mathbf{v}_n}{\partial t} &+ \left(\mathbf{v}_n\cdot\mathbf{\nabla}\right)\mathbf{v}_n \bigg] + \mathbf{\nabla}P_n =\mathbf{f}_d, \label{NeuMom}
\end{align}
\end{subequations}
which represent conservation laws of mass and momentum, respectively. The corresponding momentum equation for the ionized fluid and magnetic induction equation are
\begin{subequations}
\begin{align}
\rho_i \bigg[\frac{\partial\mathbf{v}_i}{\partial t} &+ \left(\mathbf{v}_i\cdot\mathbf{\nabla}\right)\mathbf{v}_i \bigg] + \mathbf{\nabla}P_i - \frac{1}{4\pi}\left(\mathbf{\nabla}\times\mathbf{B}\right)\times\mathbf{B}=-\mathbf{f}_d, \label{IonMom}\\
\frac{\partial\mathbf{B}}{\partial t} &+ \mathbf{\nabla}\times\left(\mathbf{B}\times\mathbf{v}_i\right) = 0.\label{induc}
\end{align}
\end{subequations}
We discuss the ion density evolution below; this must take into account ionization and recombination.

The ion-neutral drag force per unit volume is
\begin{equation}
\mathbf{f}_d = \alpha\rho_n\rho_i\left(\mathbf{v}_i-\mathbf{v}_n\right),
\label{drag}
\end{equation}
where $\left|\mathbf{v}_i-\mathbf{v}_n\right|$ is the slip speed, and $\alpha = \langle\sigma v_\mathrm{rel}\rangle/\left(\mu_n + \mu_i\right)$ is the collision coefficient with the collisional cross-section $\sigma$. The mean neutral and ion molecular weight $\mu_n$ and $\mu_i$ are applied here so the number density is $n_n = \rho_n / \mu_n$, $n_i = \rho_i / \mu_i$. For simplicity, we shall assume an isothermal equation of state, $P_n = c_{sn}^2\rho_n$, $P_i = c_{si}^2\rho_i$, and   $c_s^2 =  P/\rho = kT/\mu$. 

\subsection{Steady State One-dimensional Shock Equations}

We now consider one-dimensional solutions that are steady, $\partial/\partial t = 0$, in the shock frame. We assume the magnetic field is parallel to the shock front. The $x$ coordinate is taken to be perpendicular to $\mathbf{B}$ and the shock front. We define the compression ratio of neutral density induced by the shock:
\begin{equation} \rho_n\equiv\rho_{n,0} r_n,\end{equation}
where $r_n\rightarrow 1$ upstream, and $r_n\rightarrow const.$ downstream. Since $\rho_n v_n = const.$ from Equation~(\ref{mCon}), $ v_n = v_{n,0}/r_n$, where $v_{n,0}$ is the neutral speed far upstream.

Since magnetic flux is conserved, $v_i B = const.$ in the gas. Far upstream, $B \rightarrow B_0 = const,\ v_i \rightarrow v_{i,0} = const.$. We define the compression ratio for magnetic field such that
\begin{equation} B \equiv r_B B_0 ,\end{equation}
and $v_i = v_{i,0}/r_B$ with $r_B\rightarrow 1$ upstream and $r_B \rightarrow const.$ downstream.

For regions far from the shock there is no structure in the fluid, $\partial v/\partial x\rightarrow 0$, $\partial\rho/\partial x\rightarrow 0$, $\partial B/\partial x\rightarrow 0$. For Equations~(\ref{NeuMom}) and (\ref{IonMom}), this means $v_i = v_n$ far upstream and downstream. Therefore $v_{n,0} = v_{i,0} \equiv v_0$ far upstream, and $r_n = r_B \equiv r_f$ far downstream. The velocities of neutrals and ions are therefore given in terms of the upstream shock-frame speed $v_0$ and the compression ratios at any $x$ as
\begin{equation}
v_n = \frac{v_0}{r_n}
\label{vn}
\end{equation}
and
\begin{equation}
v_i = \frac{v_0}{r_B}.
\label{vi}
\end{equation}
To simplify the equations, we define an ion compression ratio
\begin{equation} \rho_i \equiv \rho_{i,0} r_i,\end{equation}
where $\rho_{i,0}$ is the upstream ion density, and $r_i \rightarrow 1$ upstream, $r_i \rightarrow const.$ (not necessarily equal to $r_f$) downstream, similar to $r_n$ and $r_B$.

The momentum equations can now be expressed in dimensionless form as
\begin{subequations}
\begin{align}
{\cal M}^2\frac{\partial}{\partial x}&\left(\frac{1}{r_n}\right) + \frac{\partial}{\partial x}\left(r_n\right)= \frac{\alpha\rho_{i,0}}{v_0}{\cal M}^2 r_n r_i \left(\frac{1}{r_B}-\frac{1}{r_n}\right),\label{momN}\\
\frac{\rho_{i,0}}{\rho_{0}}{\cal M}^2\frac{r_i}{r_B}&\frac{\partial}{\partial x}\left(\frac{1}{r_B}\right) + \frac{\rho_{i,0}}{\rho_{0}}\frac{\mu_n}{\mu_i}\frac{\partial}{\partial x}\left(r_i\right) + \frac{1}{\beta}\frac{\partial}{\partial x}\left(r_B^2\right)= -\frac{\alpha\rho_{i,0}}{v_0} {\cal M} ^2 r_n r_i \left(\frac{1}{r_B}-\frac{1}{r_n}\right),\label{momI}
\end{align}
\label{momEq}
\end{subequations}
in which ${\cal M}$ and $\beta$ are two dimensionless parameters defined as
\begin{equation} {\cal M}^2\equiv\left(\frac{v_0}{c_{s}}\right)^2, \ \ \ \ \ \frac{1}{\beta}\equiv\frac{B_0^2}{8\pi\rho_{0}c_{s}^2} = \frac{1}{2}\left(\frac{v_{\mathrm{A},0}}{c_{s}}\right)^2,\label{MaBeta}\end{equation}
that is, upstream values of the square of Mach number and (half of) the square of the Alfv\'{e}n Mach number of neutrals, respectively. In Equations~(\ref{momEq})$-$(\ref{MaBeta}) and subsequently, we use the shorthand notation $c_{sn}\rightarrow c_s$, $\rho_{n,0}\rightarrow\rho_0\equiv\mu_n n_0$, and $v_{\mathrm{A}n,0}\rightarrow v_{\mathrm{A},0}$. The drag force terms on the right-hand sides of Equations~(\ref{momN}) and (\ref{momI}) have equal magnitudes and opposite signs. Note that although Equations~(\ref{momN}) and (\ref{momI}) represent the case with magnetic field parallel to the shock front, the results for the case with more general geometry are qualitatively similar (see Appendix~\ref{sec:appendix} for detailed discussion).

\subsection{Governing Ordinary Differential Equation}

Typically, we have $\mu_i/\mu_n\approx 30/2.3 \approx 13$, and
\begin{equation} \alpha=\frac{\langle\sigma v_\mathrm{rel}\rangle}{\mu_i + \mu_n} \approx \frac{2\times 10^{-9}~\mathrm{cm^3 s^{-1}}}{32.3m_\mathrm{H}} = 3.7\times 10^{13}~\mathrm{cm^3 s^{-1} g^{-1}}\end{equation}
\citep{1983ApJ...264..485D}. The Mach number $\cal{M}$ is generally at least $\sim 10$, the plasma parameter is uncertain, but presumably $\beta\sim 0.01-1$, and since we are considering lightly ionized fluid, $x_{i,0}\equiv n_{i,0}/n_0$ is a very small number, $\sim10^{-6}$ (here, $n_0 = \rho_0 / \left(2.3 m_\mathrm{H}\right)$). The compression ratios $r_n$, $r_B$, and $r_i$ are dimensionless and are maximal downstream, with typical values $\sim 10$. Therefore, the last term on the left-hand side in Equation~(\ref{momI}) dominates over the other two terms. 

Retaining only the largest terms in the ion momentum equation yields
\begin{equation}
\frac{d r_B^2}{dx} = -\beta\frac{\alpha\rho_{i,0}}{v_0}{\cal M}^2 r_n r_i\left(\frac{1}{r_B}-\frac{1}{r_n}\right).\label{drB2dx}
\end{equation}
Using this result, the neutral momentum equation can be written as
\begin{equation}\frac{d}{dx}\left(\frac{{\cal M}^2}{r_n}\right) + \frac{d}{dx}\left(r_n\right) = -\frac{1}{\beta}\frac{d}{dx}\left(r_B^2\right),\label{drn+dB2}\end{equation}
or
\begin{equation}
\frac{{\cal M}^2}{r_n} + r_n + \frac{r_B^2}{\beta} = const. = {\cal M}^2 + 1 + \frac{1}{\beta},
\label{up_down}
\end{equation}
an expression of conservation of momentum of the magnetized medium. On the right-hand side of Equation~(\ref{up_down}), we have used $r_n=1=r_B$ upstream. Equations~(\ref{drB2dx}) and (\ref{drn+dB2}) represent the ``strong coupling" approximation, in which the full magnetic force on the ions is conveyed to the neutrals, i.e., 
\begin{equation}
\mathbf{f}_d = \alpha\rho_i\rho_n\left(\mathbf{v}_i - \mathbf{v}_n\right) = \frac{\left(\mathbf{\nabla}\times\mathbf{B}\right)\times\mathbf{B}} {4\pi}
\end{equation}
\citep[Equation~(27.8)]{1992phas.book.....S}.

We can solve Equation~(\ref{up_down}) to obtain
\begin{equation}
r_B = \left[1+\beta\left(r_n-1\right)\left(\frac{{\cal M}^2}{r_n}-1\right)\right]^{1/2};\label{rB}
\end{equation}
once $r_n(x)$ is known, this gives $r_B(x)$.
The compression ratio $r_f$ for both neutrals and magnetic field lines in the post-shock region is obtained by setting $r_B = r_f = r_n$ in Equation~(\ref{rB}), yielding
\begin{equation}r_f = \frac{2\beta{\cal M}^2}{1+\beta+\left[\left(1+\beta\right)^2+4\beta{\cal M}^2\right]^{1/2}}.\label{rf}\end{equation}
Note that if $\beta{\cal M}^2 \gg 1$, for a strong shock,
\begin{equation} r_f \approx \sqrt{\beta}{\cal M}= \sqrt{2}\frac{v_0}{v_{\mathrm{A},0}}. \label{rfApprox}\end{equation}
In dimensional form, this is
\begin{equation}
r_f \approx  9.8 \left(\frac{n_0}{100\mathrm{cm^{-3}}}\right)^{1/2}\left(\frac{v_0}{\mathrm{km/s}}\right)\left(\frac{B_0}{\mathrm{\mu G}}\right)^{-1}.
\label{rfApproxNum}
\end{equation}

Note that for oblique C shocks, $r_B \neq r_f$ in the post-shock region, and they both depend on the angle $\theta$ between $\mathbf{B}$ and $\mathbf{v}$. Appendix~\ref{sec:appendix} provides expressions for generalized $r_f(\theta)$ with Equation~(\ref{rfQuadra}) and $r_{B,f}(\theta)$ with Equation~(\ref{rBfApprox}).

Combining Equations~(\ref{rB}) and (\ref{drB2dx}), we obtain an ODE for $r_n$. The governing equation is
\begin{equation}
\frac{dr_n}{dx} = \frac{-D r_n r_i}{1-\frac{{\cal M}^2}{r_n^2}} \left[\frac{1}{r_n}-\frac{1}{\sqrt{1+\beta\left(r_n-1\right)\left(\frac{{\cal M}^2}{r_n}-1\right)}}\right],
\label{drndx}
\end{equation}
where
\begin{equation}
D\equiv \frac{\alpha\rho_{i,0}}{v_0}{\cal M}^2 = \frac{\alpha \mu_i}{c_s^2}x_{i,0}n_0 v_0.
\label{constD}
\end{equation}
If we use $c_s = 0.2~\mathrm{km/s}\left(T/10\mathrm{K}\right)^{1/2}$, 
\begin{equation}
D = 150~\mathrm{pc}^{-1} \left(\frac{n_0}{100~\mathrm{cm^{-3}}}\right)\left(\frac{v_0}{\mathrm{km/s}}\right)\left(\frac{x_{i,0}}{10^{-6}}\right)\left(\frac{T}{10\mathrm{K}}\right)^{-1}.
\end{equation}

\subsection{Ionization Fraction}

\begin{figure}
\epsscale{0.4}
\plotone{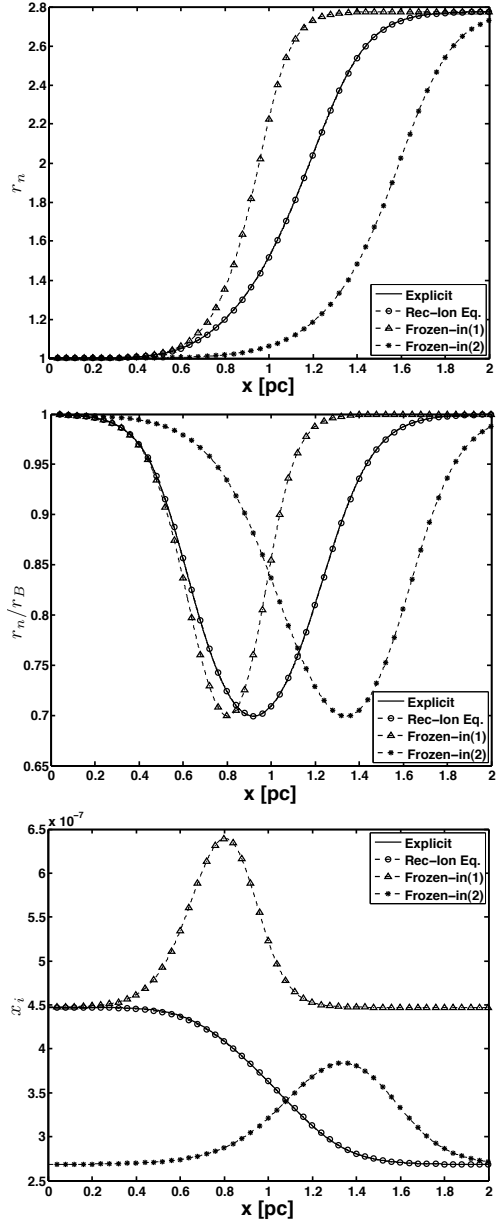}
\caption{Comparison of C shock solution with different approaches to ionization. Adopted parameters are $n_0 = 500~\mathrm{cm}^{-3}$, $v_0 = 5~\mathrm{km/s}$, $B_0=10~\mu\mathrm{G}$, and $\chi_{i0} = 10$. ``Frozen-in(1)" means upstream ionization is in equilibrium, and ``Frozen-in(2)" means downstream ionization is in equilibrium. Evidently, recombination-ionization equilibrium (open circles) is an excellent approximation to the exact solution (solid curve).}
\label{compStruc}
\end{figure}

To solve the ODE in Equation~(\ref{drndx}), we need a relation between $r_n$ and $r_i$. In the dense interstellar medium, the main source of neutral ionization is cosmic rays, while ions may recombine in the gas phase, or on dust grains. The evolution of ion number density can be written as
\begin{equation}\frac{dn_i}{dt}=\zeta_\mathrm{CR} n_n - \alpha_\mathrm{gas} n_i^2 - \alpha_\mathrm{grain} n_i n_n.\end{equation}
Comparing the orders of magnitude of the three coefficients, $\zeta_\mathrm{CR}\sim 10^{-17}-10^{-16}~\mathrm{s^{-1}}$ for cosmic ray ionization \citep{1992phas.book.....S, 1983ApJ...264..485D}, $\alpha_\mathrm{gas}\sim 10^{-7}-10^{-5}~\mathrm{cm^3 s^{-1}}$ \citep[][Table 4.11]{2005pcim.book.....T}, and $\alpha_\mathrm{grain}\sim 10^{-15}~\mathrm{cm^3 s^{-1}}$ when $T\sim 10~\mathrm{K}$ \citep{2001ApJ...563..842W}. In moderate-density clouds $n_i/n_n\sim 10^{-5}-10^{-7}$ and $n_n\sim 10^2-10^3~\mathrm{cm^{-3}}$, so we can drop the grain surface recombination term. The ion balance equation becomes
\begin{equation}
\frac{\partial n_i}{\partial t} + \mathbf{\nabla}\cdot(n_i \mathbf{v}_i) \approx \zeta_\mathrm{CR} n_n - \alpha_\mathrm{gas} n_i^2.\label{equilibrium}
\end{equation}

\subsubsection{Recombination-Ionization Equilibrium}
In solving Equation~(\ref{equilibrium}), one possible approximation is to assume ionization-recombination equilibrium everywhere. In this case, $\zeta_\mathrm{CR} n_n \approx \alpha_\mathrm{gas} n_i^2$, so that
\begin{equation}n_i = \sqrt{\frac{\zeta_\mathrm{CR}}{\alpha_\mathrm{gas}}} n_n^{1/2} \equiv 10^{-6} \chi_{i0} n_n^{1/2},\label{rec-ion}\end{equation}
for
\begin{equation}
\chi_{i0} \equiv 10^6 \times \sqrt{\frac{\zeta_\mathrm{CR}}{\alpha_\mathrm{gas}}},
\label{chiDef}
\end{equation}
where the coefficient $\chi_{i0}\sim 1-20$ \citep{2010ApJ...720.1612M}. 

If we adopt Equation~(\ref{rec-ion}), then $r_i = r_n^{1/2}$, and the governing ODE becomes
\begin{equation}\frac{dr_n}{dx} = \frac{-D r_n^{3/2}}{1-\frac{{\cal M}^2}{r_n^2}} \left(\frac{1}{r_n}-\frac{1}{r_B}\right),
\label{govEq}
\end{equation}
where $r_B$ is given in terms of $r_n$ by Equation~(\ref{rB}).

\subsubsection{Frozen-in Magnetic Field}

Another approach to Equation~(\ref{equilibrium}) is the so-called frozen-in condition \citep[e.g.,][]{1990MNRAS.246...98W}, which has been applied widely. In this approximation, ionizations and recombinations are neglected, so that for a steady flow, $n_i v_i = const.$, which implies $r_i = r_B$. This corresponds to a ``frozen-in field": the compression ratio of the magnetic field is the same as the ion flow. The governing ODE then becomes (using Equation~(\ref{rB}))
\begin{equation}
\frac{dr_n}{dx} = \frac{-D r_n}{1-\frac{{\cal M}^2}{r_n^2}} \left[\frac{\sqrt{1+\beta\left(r_n-1\right)\left(\frac{{\cal M}^2}{r_n}-1\right)}}{r_n}-1\right].
\end{equation}

One thing worth noting here is that in the frozen-in approximation, the ionization fraction in the post-shock region will be the same as in the upstream region. Since Equation~(\ref{rec-ion}) must hold far upstream and far downstream, we must choose whether to set $x_{i,0}$ based on $n_0$ or $r_f n_0$.

\subsubsection{Explicit Solution}

We can also retain all terms in the ionization-recombination equation in our numerical integration. Using $n_i = n_0 r_n x_i$, Equation~(\ref{equilibrium}) in steady state, for one dimension, yields
\begin{equation}
\frac{d x_i}{dx} = \frac{\zeta_\mathrm{CR}}{v_0} r_B - \frac{\alpha_\mathrm{gas} n_0}{v_0} x_i^2 r_n r_B + x_i \frac{d}{dx}\left[\ln\left(\frac{r_B}{r_n}\right)\right],
\label{dxidx}
\end{equation}
Here, $r_B$ is given in terms of $r_n$ by Equation~(\ref{rB}). By integrating Equations~(\ref{dxidx}) and (\ref{drndx}) together, we can calculate the explicit solution for the steady C shock system.

\subsubsection{Comparison of Ionization Treatments}

To compare ionization-recombination equilibrium and the explicit solution, we choose just an upstream value $\chi_{i0}$. For the frozen-in field case, we must also choose whether our solution will have the same upstream ionization fraction as the equilibrium case, or the same downstream value as the equilibrium case. Therefore there are four different cases for us to compare.

An example comparing the shock solutions for the four different ionization choices is shown in Fig.~\ref{compStruc}. Evidently, the approximation of ionization-recombination equilibrium yields a solution very close to the explicit solution. We have found that this is true for the full range of parameters of interest, $n_0\sim 10^2$ to $10^3~\mathrm{cm^{-3}}$, $v_0\sim 1$ to $10~\mathrm{km/s}$, $B_0\sim 1$ to $15~\mathrm{\mu G}$, $\chi_{i0} \sim 1$ to $10$. Henceforth, we shall adopt ionization-recombination equilibrium and use $n_i\propto n_n^{1/2}$ so that $r_i = r_n^{1/2}$, and Equation~(\ref{govEq}) governs steady C shocks.

\section{Steady C Shock Thickness}
\label{sec: thickness}

For any given set of parameters $n_0$, $v_0$, $B_0$, and $\chi_{i0}$, Equation~(\ref{govEq}) can be integrated to obtain a steady C shock solution. However, it is also useful to obtain estimates of the dependence of the C shock thickness on the basic flow parameters. This parameterization is potentially useful in diagnosing magnetic field strengths from observations. In addition, it provides a helpful guide to assessing the scales at which ambipolar diffusion becomes important in GMCs dominated by strong turbulence. If, by appropriate simplifications we can integrate the governing ODE of Equation~(\ref{govEq}) analytically, we can obtain an approximate expression for the shock thickness as a function of $n_0$, $v_0$, $B_0$, and $\chi_{i0}$. Note that, since the governing equations for oblique shocks are qualitatively similar to the simplified case applied here, the oblique shock thickness can be approached using the same methods discussed in this section (see Appendix~\ref{sec:appendix}).

\subsection{Exact Solution}
\label{sec:exact}

\begin{figure}
\epsscale{0.5}
\plotone{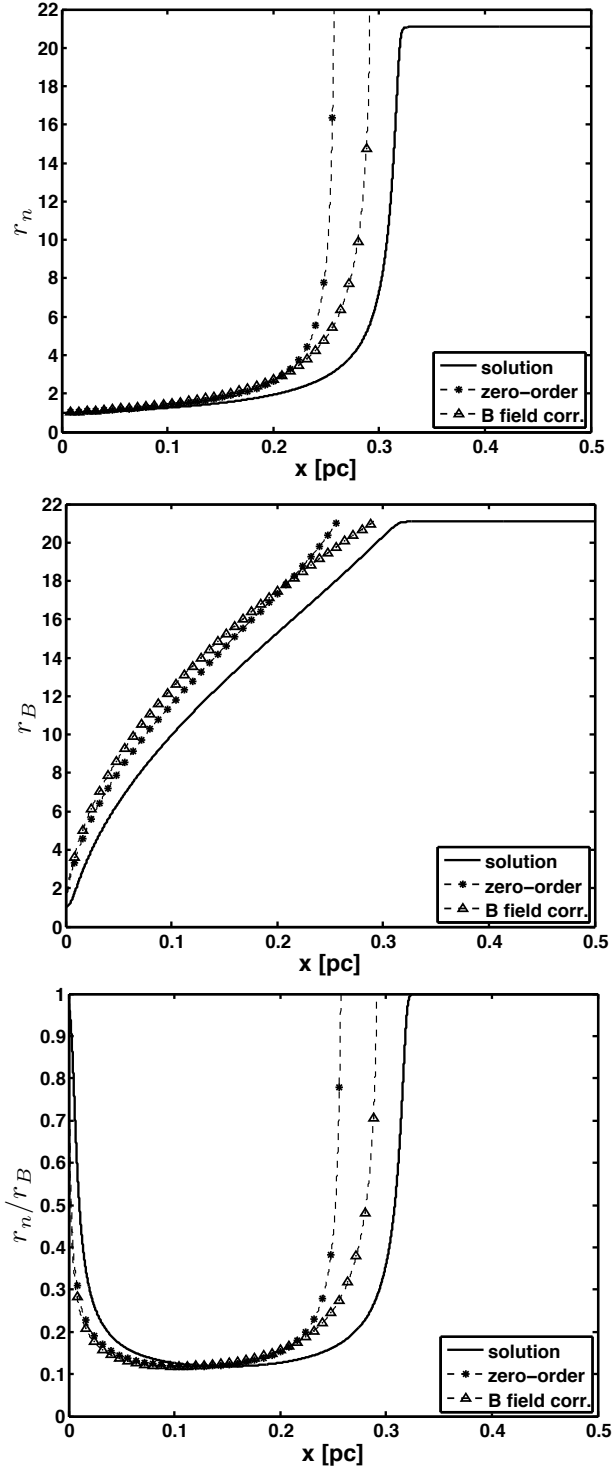}
\caption{Exact C shock solution (solid) compared to the ``zeroeth-order" estimate of Equation~(\ref{0thApprox}) (circles) and an improved approximation given by Equation~(\ref{rnAna2}) (triangles), for parameters $n_0 = 500 ~\mathrm{cm^{-3}}$, $v_0 = 5 ~\mathrm{km/s}$, $B_0=5~\mu \mathrm{G}$, and $\chi_{i0}=10$.}
\label{ana_thick1}
\end{figure}

\begin{figure}
\epsscale{0.5}
\plotone{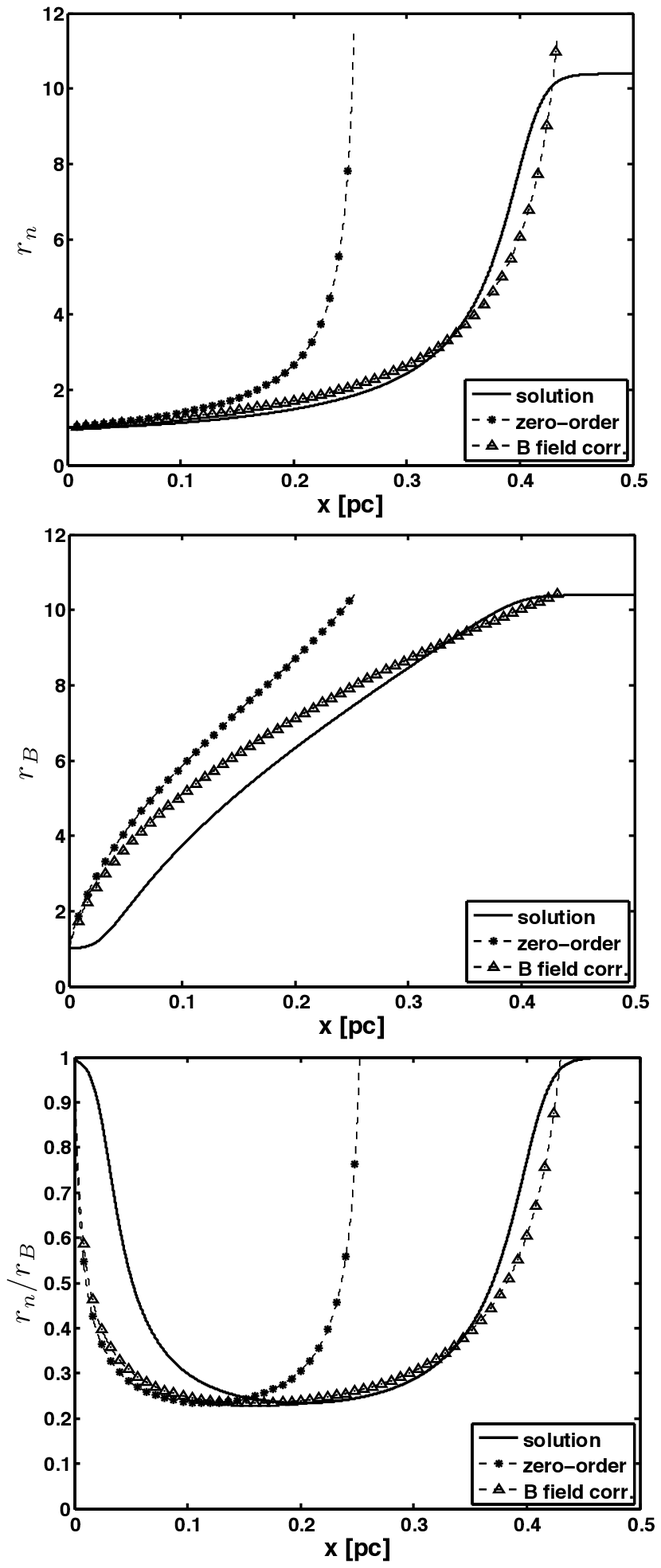}
\caption{Same as Fig.~\ref{ana_thick1}, for $n_0 = 500 ~\mathrm{cm^{-3}}$, $v_0 = 5 ~\mathrm{km/s}$, $B_0=10~\mu \mathrm{G}$, and $\chi_{i0}=10$.}
\label{ana_thick2}
\end{figure}

From numerical integrations of Equation~(\ref{govEq}) with a range of parameters, we have found that $r_n/r_B$ drops very quickly at the beginning, becomes flat in the central region, then increases rapidly near the other edge of the shock (see bottom panels of Fig.~\ref{ana_thick1} and \ref{ana_thick2}). This behavior can be used to define the thickness of C-type shocks. Since the minimum of $r_n/r_B$ depends on the parameters (see Equation~(\ref{rndrBmin}) below), we should ensure that our thickness definition is insensitive to this value. Based on these considerations, we adopt the following definition of shock thickness for exact numerical solutions:
\begin{equation}x_s\equiv x\Big|_{r_n/r_B=0.95},\ \ x_f\equiv x\Big|_{r_n/r_B=0.95},\ \ x_f > x_s;\ \ \Rightarrow\  \mathrm{shock\ thickness}\ \ L_\mathrm{exact} \equiv \big|x_f - x_s\big|.\label{thickDef}\end{equation}
Note that for some weak shocks, $r_n/r_B$ is always larger than $0.95$. Therefore this definition also provides limitations in the parameter space to exclude shocks which are not strong and thus do not satisfy our strong shock analysis. 

We have integrated the shock ODE for a range of parameters, and computed the shock thickness according to the definition in Equation~(\ref{thickDef}). This is the dataset of exact solutions of C shock thickness over a parameter grid with $10$ values of $n_0$ equally spaced between $10^2$ and $10^3$~$\mathrm{cm^{-3}}$, $14$ values of $v_0$ equally spaced between $2$ and $15$~$\mathrm{km/s}$, $14$ values of $B_0$ equally spaced between $2$ and $15$~$\mu\mathrm{G}$, and $11$ values of $\chi_{i0}$ equally spaced between $1$ and $21$. The range of C shock thickness is $0.1$ to $20$~pc in this parameter range. Note that all parts of this parameter space are not necessarily astronomically realistic. For example, low $n_0$ and high $v_0$ is unlikely to have low $\chi_{i0}$, so very large C shock thickness is not likely to be found.

Also note that even for C shock thickness $\sim 1$~pc, in a real molecular cloud all the parameters are likely to vary within this length scale, instead of staying constant as in our models. However, our solutions still provide a useful guide to approximate shock thicknesses for parameters within a given range.

\subsection{Zeroeth-order Approximation}

We consider the relative magnitudes of the terms in Equations~(\ref{rB}) and (\ref{govEq}). First, since $c_s \sim0.2~\mathrm{km/s}$ whereas $v_0 \gtrsim 1~\mathrm{km/s}$, in general ${\cal M}^2$ is a very large number, and typically ${\cal M}^2 \gg r_n^2$ (see Equation~(\ref{rfApprox})). Also, from Fig.~\ref{compStruc}, the ratio $r_n/r_B$ is small in much of the shock region. If we let $r_n/r_B \ll 1$, a ``zeroeth-order" approximation to Equation~(\ref{govEq}) is
\begin{equation}
\frac{dr_n}{dx}\approx \frac{D r_n^{5/2}}{{\cal M}^2}.
\end{equation}
This can be integrated analytically to yield 
\begin{equation}
r_n (x) = \left(1-\frac{3}{2}\frac{Dx}{{\cal M}^2}\right)^{-2/3}.
\label{0thApprox}
\end{equation}
This ``zeroeth-order" approximation to the shock structure using Equation~(\ref{0thApprox}) is shown in Fig.~\ref{ana_thick1} and \ref{ana_thick2} for two parameter sets, in comparison to the exact solution. The zeroeth-order shock thickness $L_\mathrm{zeroeth}$ is defined as $x$ such that $r_n\rightarrow r_f$ in Equation~(\ref{0thApprox}), giving
\begin{equation}L_\mathrm{zeroeth} = \frac{2}{3}\frac{{\cal M}^2}{D}\left(1-r_f^{-3/2}\right)\approx\frac{2}{3}\frac{{\cal M}^2}{D},\label{thickApprox}\end{equation}
where the second approximation assumes a strong shock, $r_f \gg 1$. 

Substituting Equation~(\ref{constD}) for $D$ in Equation~(\ref{thickApprox}), we obtain a thickness estimate in terms of physical parameters
\begin{equation}
L_\mathrm{zeroeth} =\frac{2}{3}\frac{v_0}{\alpha\rho_{i,0}} \propto \frac{v_0}{\chi_{i0}n_0^{1/2}},\label{thick_ana}
\end{equation}
or in dimensional form,
\begin{equation} 
L_\mathrm{zeroeth} \approx 0.12~\mathrm{pc} \times \left(\frac{n_0}{100\mathrm{cm^{-3}}}\right)^{-1/2}\left(\frac{v_0}{\mathrm{km/s}}\right)\left(\frac{\chi_{i0}}{10}\right)^{-1} .
\label{thickApprox0}
\end{equation}
Thus, the shock thickness increases with higher upstream velocity, and decreases with higher upstream neutral density and ionization fraction. In this ``zeroeth-order" approximation the shock thickness does not depend on the upstream magnetic field strength. From the examples shown in Fig.~\ref{ana_thick1} and \ref{ana_thick2}, we can see that although the zeroeth-order solution follows the general behavior of C shocks, it is not accurate for strongly-magnetized cases (Fig.~\ref{ana_thick2}). Compared with the dataset of exact solutions discussed in previous section, the RMS value of $(L_\mathrm{exact}-L_\mathrm{zeroeth})/L_\mathrm{exact}$ is $0.355$, and the range of $(L_\mathrm{exact}-L_\mathrm{zeroeth})/L_\mathrm{exact}$ is $-0.8$ to $0.28$.

The dependence on the velocity, ion density, and collision coefficient in Equation~(\ref{thick_ana}) can be understood in terms of the drag force between ions and neutrals. The total momentum flux in neutrals entering the shock is $\rho_0{v_0}^2$. The mean drag force per volume is $\sim\alpha\rho_0\rho_{i,0}v_0$. The ratio of these quantities, which is the characteristic distance over which momentum exchange takes place, is 
\begin{equation} L\sim\frac{\rho_0{v_0}^2}{\alpha\rho_0\rho_{i,0}v_0}\sim\frac{v_0}{\alpha\rho_{i,0}}\propto v_0 n_0^{-1/2}\chi_{i0}^{-1}.\end{equation}
This dependence is similar to Equation~(3.12) in \cite{1993ARA&A..31..373D} if the Alfv\'{e}n speed in the fluid is similar to the upstream velocity, $v_\mathrm{A}\sim v_0$. Although they obtained an estimate using different assumptions and approximations, the basic idea that the momentum transfer rate determines the shock thickness is similar.

\subsection{Magnetic Field Influence}
\label{sec:ThickMag}

To obtain a more accurate estimate of the C shock thickness, we return to the differential equation~(\ref{momN}) for neutral momentum flux, making use of Equation~(\ref{constD}) and the ionization equilibrium condition $r_i = r_n^{1/2}$,
\begin{equation}
\frac{d}{dx}\left(r_n+\frac{{\cal M}^2}{r_n}\right) = -Dr_n^{3/2}\left(\frac{1}{r_n}-\frac{1}{r_B}\right) = -Dr_n^{1/2}\left(1-\frac{r_n}{r_B}\right).
\label{drnThickness}
\end{equation}
We integrate this equation, using constant values on the right-hand-side
\begin{equation}
\left\langle r_n^{1/2} \right\rangle \rightarrow \frac{1+r_f^{1/2}}{2}\approx \frac{\sqrt{r_f}}{2},
\ \ \ \ 
\left\langle 1-\frac{r_n}{r_B}\right\rangle \rightarrow \left(\frac{1-\left(r_n/r_B\right)_\mathrm{min}}{2}\right),
\end{equation}
where the minimum value of $r_n/r_B$ can be derived explicitly from Equation~(\ref{rB}) as
\begin{equation}
\frac{r_n}{r_B}\bigg|_\mathrm{min} = \frac{3\sqrt{3}}{2\sqrt{\beta}{\cal M}}.
\label{rndrBmin}
\end{equation}
This yields a quadratic for $r_n$ as a function of $x$:
\begin{equation}
r_n^2 - \left({\cal M}^2 + 1 - D\left\langle r_n^{1/2} \right\rangle \left\langle 1-\frac{r_n}{r_B}\right\rangle x \right) r_n + {\cal M}^2 = 0. \label{rnAna2}
\end{equation}
Solving Equation~(\ref{rnAna2}) for $r_n(x)$ gives us another analytical approximation of the shock structure. When compared with the explicit solution and the zeroth-order approximation in Fig.~\ref{ana_thick1} and~\ref{ana_thick2}, we can see that this correction is necessary only when the background magnetic field is strong (Fig.~\ref{ana_thick2}).

For Equation~(\ref{rnAna2}) the magnetically-corrected estimate of the shock thickness ($x=L_\mathrm{est}$ such that $r_n=r_f$) can be written as
\begin{equation}
L_\mathrm{est} = \frac{\left({\cal M}^2-r_f\right) \left(r_f -1\right)}{D \left\langle r_n^{1/2} \right\rangle \left\langle 1-r_n/r_B\right\rangle r_f}.
\end{equation}
Assuming ${\cal M}^2\gg r_f\gg 1$ and $(r_n/r_B)_\mathrm{min} \ll 1$, and using Equation~(\ref{constD}), we have
\begin{equation}
L_\mathrm{est} \approx \frac{4{\cal M}^2}{D r_f^{1/2}} = \frac{4v_0}{\alpha\rho_{i,0}r_f^{1/2}}.
\label{anaL1}
\end{equation}
Note that a similar result can be obtained for the generalized case with an oblique C shock (Equation~(\ref{ObLapprox})). See Appendix~\ref{sec:appendix} for detailed discussion.

Taking the strong-compression limit $r_f \approx \sqrt{2} v_0/v_{\mathrm{A},0}$ of Equation~(\ref{rfApprox}), we have
\begin{equation}
L_\mathrm{est} = \frac{2^{7/4} v_0^{1/2} v_{\mathrm{A},0}^{1/2}}{\alpha\rho_{i,0}} \propto n_0^{-3/4}v_0^{1/2}B_0^{1/2}\chi_{i0}^{-1},
\label{thickAna_B1}
\end{equation}
or in dimensional form
\begin{equation}
L_\mathrm{est} = 0.22~\mathrm{pc}\times\left(\frac{n_0}{100\mathrm{cm^{-3}}}\right)^{-0.75}\left(\frac{v_0}{\mathrm{km/s}}\right)^{0.5}\left(\frac{B_0}{\mathrm{\mu G}}\right)^{0.5}\left(\frac{\chi_{i0}}{10}\right)^{-1}.\label{thickAna_B2}
\end{equation}
Compared with Equation~(\ref{thick_ana}), the shock thickness still depends positively on inflow velocity and negatively on upstream density and ionization fraction, but now a dependence on the magnetic field enters as well. Compared with the dataset of exact solutions discussed above, the RMS value of $(L_\mathrm{exact}-L_\mathrm{est})/L_\mathrm{exact}$ is $0.13$, and the range of $(L_\mathrm{exact}-L_\mathrm{est})/L_\mathrm{exact}$ is $-0.21$ to $0.26$. \cite{1990MNRAS.246...98W} and \cite{2006ApJ...653.1280L} find $L_\mathrm{shock}\sim \sqrt{2} v_{\mathrm{A},0}/\left(\alpha\rho_{i.0}\right)$ in the case where ions are frozen in; this is smaller than Equation~(\ref{thickAna_B1}) by a factor $2^{-5/4}\left(v_{\mathrm{A},0}/v_0\right)^{1/2}$.

\subsection{Numerical Approach}

Using the dataset of exact solutions discussed in Section~\ref{sec:exact}, we construct a simultaneous linear fit for $\log L_\mathrm{exact}$ to $\log n_0$, $\log B_0$, $\log v_0$, and $\log\chi_{i0}$. We find
\begin{equation} 
L_\mathrm{fit} = 0.21~\mathrm{pc}\times\left(\frac{n_0}{100\mathrm{cm^{-3}}}\right)^{-0.73}\left(\frac{v_0}{\mathrm{km/s}}\right)^{0.54}\left(\frac{B_0}{\mathrm{\mu G}}\right)^{0.46}\left(\frac{\chi_{i0}}{10}\right)^{-1}.\label{thick_num}
\end{equation}
Over the parameter grid, the RMS value of $(L_\mathrm{exact}-L_\mathrm{fit})/L_\mathrm{exact}$ is $0.08$, and the range of $(L_\mathrm{exact}-L_\mathrm{fit})/L_\mathrm{exact}$ is $-0.29$ to $0.22$.

The result in Equation~(\ref{thick_num}) agrees with our expectation that the shock thickness depends on the magnetic field. Also, the dependences on all parameters are extremely close to Equation~(\ref{thickAna_B2}). Table~\ref{thickness} lists a set of model parameters (to be used in time-dependent simulations) and the C shock thickness based on the analytic estimate in Equation~(\ref{anaL1}) and the multivariate fit in Equation~(\ref{thick_num}), in comparison with the results from explicit integration of the ODE (Equation~(\ref{govEq})). Both approaches are useful to estimate the shock thickness.

\begin{deluxetable}{cccccccc}
\tabletypesize{\small}
\tablecolumns{8}
\tablecaption{Steady C shock Thickness Comparison \label{thickness}}
\tablewidth{0pt}
\tablehead{
\colhead{Model} & \colhead{$n_0$}   & \colhead{$v_0$} & \colhead{$B_0$} & \colhead{$\chi_{i0}$} 
&\multicolumn{3}{c}{$L_\mathrm{shock}$ (pc)} \\

\colhead{}& \colhead{} & \colhead{}   & \colhead{} &\colhead{}
& \colhead{exact} & \colhead{est.}  & \colhead{fit} \\

\colhead{}& \colhead{$\left(\mathrm{cm^{-3}}\right)$}   & \colhead{$\left(\mathrm{km/s}\right)$}  &   \colhead{$\left(\mu\mathrm{G}\right)$} & \colhead{} 
& \colhead{eq.~(\ref{govEq})} & \colhead{eq.~(\ref{anaL1})}  & \colhead{eq.~(\ref{thick_num})} 
}
\startdata
N01  &   100  &  5 & 10 &  5  & 3.03 & 3.15 & 2.89\\
N03  &   300  &  5 & 10 &  5  & 1.20 & 1.38 & 1.30\\
N05  &   500  &  5 & 10 &  5  & 0.82 & 0.94 & 0.89\\
N08  &   800  &  5 & 10 &  5  & 0.58 & 0.66 & 0.63\\
N10  & 1000  &  5 & 10 &  5  & 0.50 & 0.56 & 0.54\\
V04  &   200  &  4 & 10 &  5  & 1.41 &  1.68 & 1.54\\
V06  &   200  &  6 & 10 &  5  & 1.55 & 2.05 & 1.92\\
V08  &   200  &  8 & 10 &  5  & 1.79 & 2.37 & 2.24\\
V10  &   200  &10 & 10 &  5  & 2.08 & 2.65 & 2.53\\
V12  &   200  &12 & 10 &  5  & 2.38 & 2.90 & 2.79\\
B02  &   200  &  5 &   2 &  5  & 0.92 & 0.84 & 0.83\\
B04  &   200  &  5 &   4 &  5  & 1.08 & 1.18 & 1.14\\
B06  &   200  &  5 &   6 &  5  & 1.26 & 1.45 & 1.38\\
B08  &   200  &  5 &   8 &  5  & 1.46 & 1.68 & 1.57\\
B10  &   200  &  5 & 10 &  5  & 1.66 & 1.87 & 1.74\\
B12  &   200  &  5 & 12 &  5  & 1.89 & 2.05 & 1.89\\
B14  &   200  &  5 & 14 &  5  & 2.12 & 2.22 & 2.03\\
X01  &   200  &  5 & 10 &  1  & 8.32 & 9.37 & 8.71\\
X06  &   200  &  5 & 10 &  6  & 1.39 & 1.56 & 1.45\\
X10  &   200  &  5 & 10 & 10  & 0.83 & 0.94 & 0.87\\
X15  &   200  &  5 & 10 & 15  & 0.55 & 0.62 & 0.58\\
X20  &   200  &  5 & 10 & 20  & 0.42 & 0.47 & 0.44\\
\enddata
\end{deluxetable}

\section{C Shock Formation}
\label{sec: formation}

\subsection{Numerical Algorithm for Ambipolar Diffusion}

To investigate how C shocks develop in time, we use a modified version of the numerical MHD code, \textit{Athena} \citep{2008ApJS..178..137S}. \textit{Athena} employs a single-step, directionally unsplit Godunov scheme to obtain conservative, second-order accurate solutions of the ideal MHD equations \citep{2005JCoPh.205..509G}.

In the strong coupling limit, the drag force $\mathbf{f}_d=\alpha\rho_i\rho_n\left(\mathbf{v}_i-\mathbf{v}_n\right)$ is equal to the Lorentz force
$\mathbf{f}_\mathrm{L} = \left[\left(\mathbf{\nabla}\times\mathbf{B}\right)\times\mathbf{B}\right]/\left(4\pi\right)$.
The momentum equation for neutrals is thus identical to that in the ideal MHD limit. The mass conservation equation for neutrals is also the same as for ideal MHD. In this approximation, $\mathbf{v}_i = \mathbf{v}_n + \left[\left(\mathbf{\nabla}\times\mathbf{B}\right)\times\mathbf{B}\right]/\left(4\pi\alpha\rho_i\rho_n\right)$, so that the induction equation~(\ref{induc}) becomes
\begin{equation}\frac{\partial\mathbf{B}}{\partial t} - \mathbf{\nabla}\times\left(\mathbf{v_n}\times\mathbf{B}\right) = \mathbf{\nabla}\times\left[\frac{\left(\left(\mathbf{\nabla}\times\mathbf{B}\right)\times\mathbf{B}\right)\times\mathbf{B}}{4\pi\rho_i\rho_n\alpha}\right].\label{newInduc}\end{equation}
With $\mathbf{v}_d = \mathbf{v}_i-\mathbf{v}_n$ the drift velocity between ions and neutrals, we can write the correction term in Equation~(\ref{newInduc}) in terms of a ``drift" EMF,
\begin{equation}\mathbf{\mathcal{E}}_d = \mathbf{v}_d\times\mathbf{B} = \frac{\left[\left(\mathbf{\nabla}\times\mathbf{B}\right)\times\mathbf{B}\right]\times\mathbf{B}}{4\pi\rho_i\rho_n\alpha}.\end{equation}

In our simplified 1-D problem, $\mathbf{B}=B_y\hat{\mathbf{y}}$, $\mathbf{v}=v_x\hat{\mathbf{x}}$, and the discretized magnetic field corrected by ambipolar diffusion before each step, at interface position $i\Delta x$ and time $n\Delta t$, is
\[B_y\big|_i^{n+1} = B_y\big|_i^n + \frac{\Delta t}{\Delta x}\left(\mathcal{E}_{d,z}\big|_{i+\frac{1}{2}}^n - \mathcal{E}_{d,z}\big|_{i-\frac{1}{2}}^n \right),\]
and
\[ \mathcal{E}_{d,z}\big|_{i+\frac{1}{2}}^n = \frac{1}{4\pi\alpha}\left(\frac{B_y\big|_{i+1}^n - B_y\big|_{i}^n}{\Delta x}\right)\frac{B_y^2}{\rho_i\rho_n}\bigg|_{i+\frac{1}{2}}^n.\]
This term is implemented in {\it Athena} as an operator-split update to the magnetic field. The mesh resolution is set to be $0.01$~pc.

In setting the timestep, we implement the super-timestepping approach as described by \cite{2009ApJS..181..413C}, choosing the factor $\nu=0.2$, and taking the ambipolar diffusion timestep 
\begin{equation}
\Delta t_\mathrm{AD} = 2\pi\alpha\left(\mathrm{CFL\ number}\right)\left(\Delta x\right) ^2\cdot \mathrm{min}\left[\frac{\rho_n\rho_i}{B_y^2}\right],
\end{equation}
where the CFL number is set to be $0.8$ in all simulations.

For figures presenting numerical results, $n_n\rightarrow n$ and $\mathbf{v}_n\rightarrow\mathbf{v}$.

\subsection{Convergent Flow Test}
\subsubsection{Simple Convergent Flow Test}

\begin{figure}
\epsscale{0.8}
\plotone{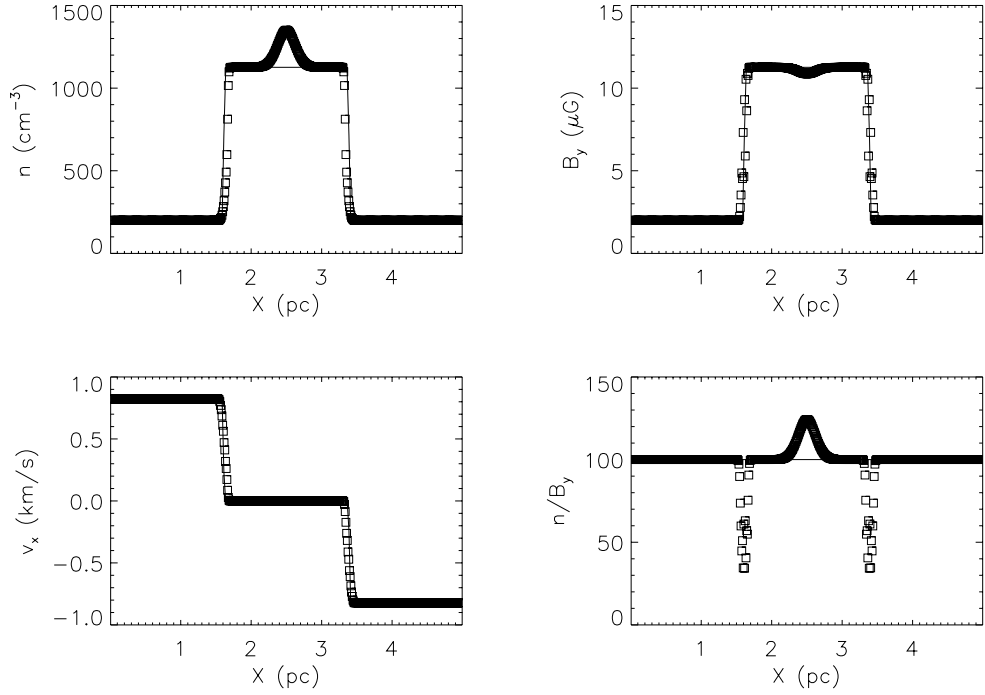}
\caption{Transient C shock structure (squares) compared with ideal MHD shock (thin lines), generated from convergent flow, for parameters $n_0 = 200~\mathrm{cm^{-3}}$, $v_0 = 1~\mathrm{km/s}$, $B_0 = 2~\mathrm{\mu G}$, and $\chi_{i0} = 10$. The central peak in density $n$ and mass-to-flux ratio $n/B_y$ is a signature of early C shock structure.}
\label{cvg1}
\end{figure}

One way to produce shocks in a numerical simulation is to use a simple convergent flow, in which the initial conditions are
\begin{equation}n = const.,\ \ \ \ B_y = const.,\ \ \ \ v_x = \left\{\begin{array}{ll} v_\mathrm{inflow},  &\mbox{left half} \\
                                                                                                                -v_\mathrm{inflow}, &\mbox{right half.}\end{array}\right.\label{cflIC}\end{equation}
This will evolve to a dense post-shock region in the center, with outward-propagating reverse C-type structures at the left and right (Fig.~\ref{cvg1}).

The C shock structures seen at $x\approx 1.5$ and $x\approx 3.5$ in Fig.~\ref{cvg1} are the same as the steady solutions obtained by integration of Equation~(\ref{govEq}), as confirmed by comparing the detailed profiles (not shown). Note that the mass-to-flux ratio $n/B_y$ is analogous to $r_n/r_B$ (except not normalized by upstream values). The dips in $n/B_y$ at the C shock locations correspond to the ``well" in $r_n/r_B$ seen in Fig.~\ref{ana_thick1} and~\ref{ana_thick2}. However, the central peaks in both neutral density $n$ and the mass-to-flux ratio $n/B_y$ in Fig.~\ref{cvg1} are not a feature of steady C shocks. As the solution shown in Fig.~\ref{cvg1} evolves further in time, these peaks disappear. Thus, these peaks are a signature of transient C shock development, as we discuss further below.

\subsubsection{Colliding Clouds}

The initial conditions for the simple converging flow are somewhat artificial, in that only the velocity is discontinuous. Thus, we would like to test whether shocks formed under more realistic conditions also show the transient peaks in $n$ and $n/B_y$ described above. 

We consider the collision of two idealized clumps inside a large molecular cloud. We suppose that the two clumps are both denser than their surrounding, but the mass-to-flux ratios are the same throughout the whole cloud. We imagine that the large-scale turbulence in the molecular cloud imposes velocities such that the two dense clumps collide with each other, producing a shock. We simulate the scenario described, setting the background density to be $5\%$ of the value in the dense clumps, and the initial velocity of this gas to be zero. We focus just on the collision region, so that the right and left sides of the domain are set to ``clump" conditions, as in Equation~(\ref{cflIC}). 

\begin{figure}
\epsscale{0.8}
\plotone{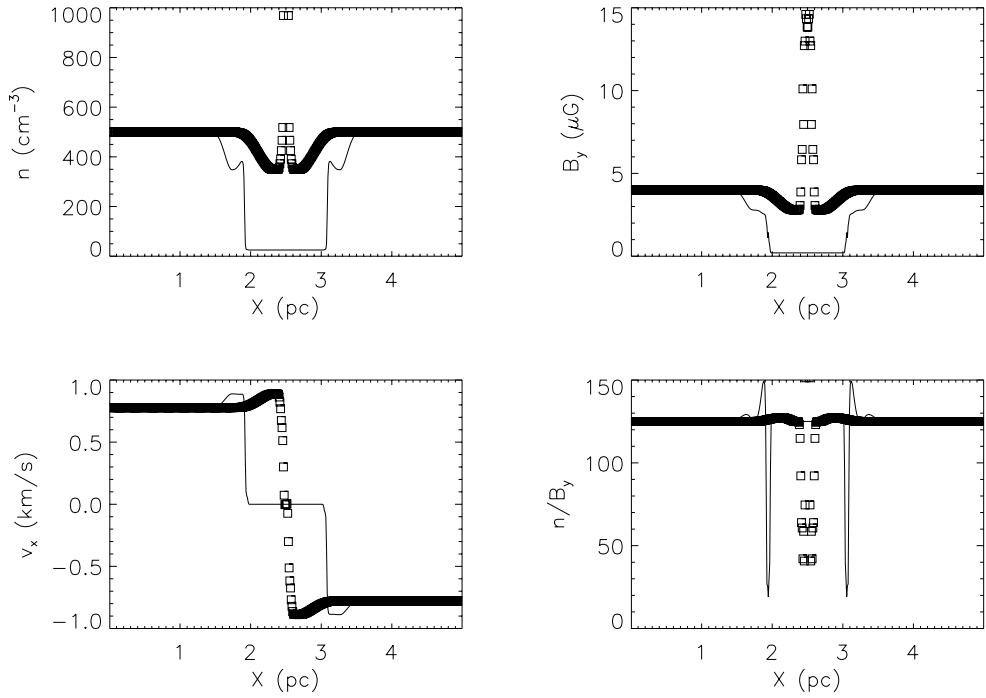}
\caption{Two dense clumps collide with each other and produce a shock (squares). Conditions at a time $0.88$~Myr prior to the collision are shown as thin lines for comparison. Parameters are $n_0 = 500~\mathrm{cm^{-3}}$, $v_0 = 1~\mathrm{km/s}$, $B_0 = 4~\mathrm{\mu G}$, and $\chi_{i0} = 10$.}
\label{CC3}
\end{figure}

\begin{figure}
\epsscale{0.8}
\plotone{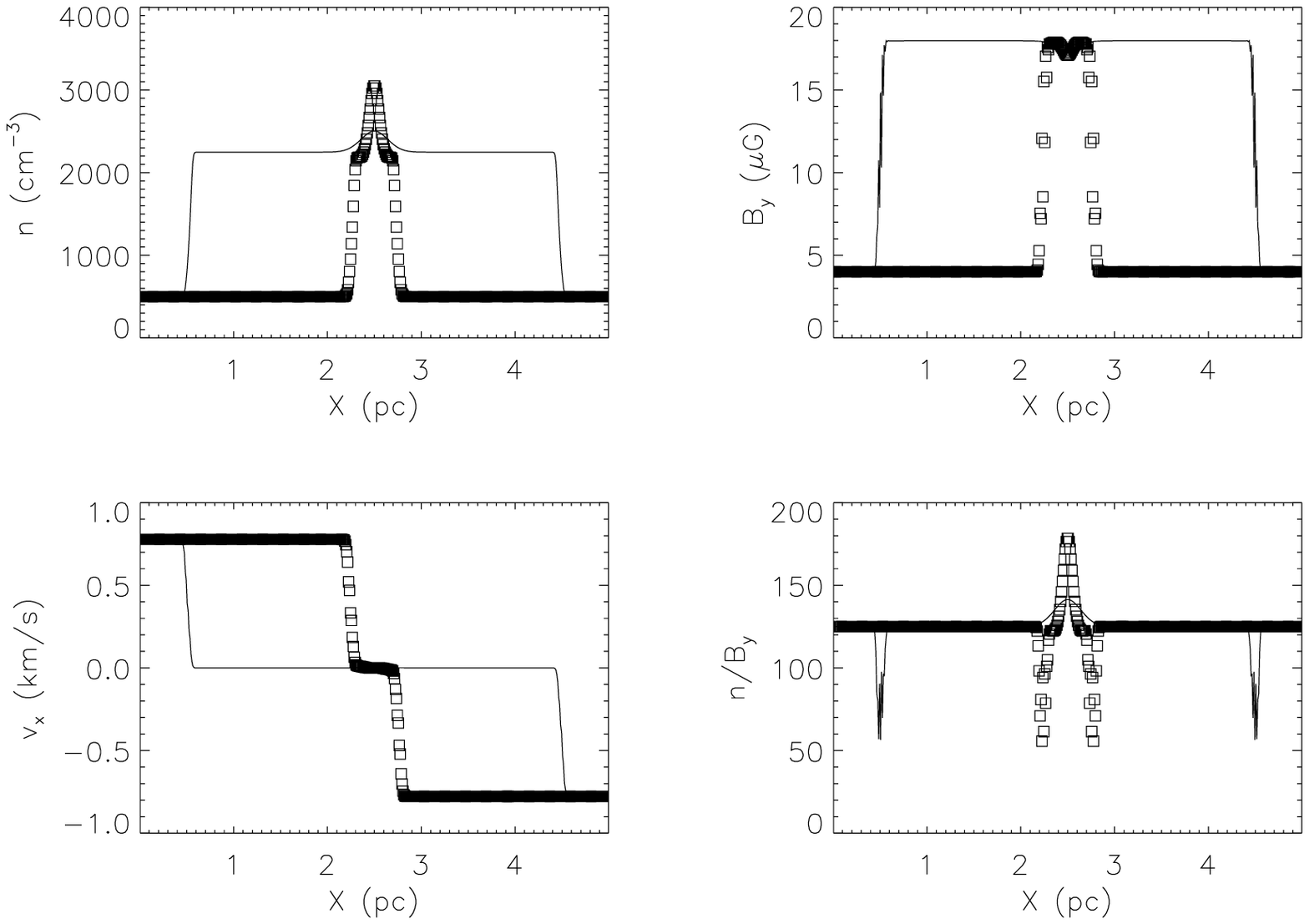}
\caption{Shock structure generated from the colliding clumps (squares), at a time $1.08$~Myr after the stage shown in Fig.~\ref{CC3}. Note that the central peaks in $n$ and $n/B_y$ are qualitatively similar to those in Fig.~\ref{cvg1}. These central peaks will then expand and smooth out (thin lines show solution after an additional $7.67$~Myr). }
\label{CC4}
\epsscale{0.8}
\plotone{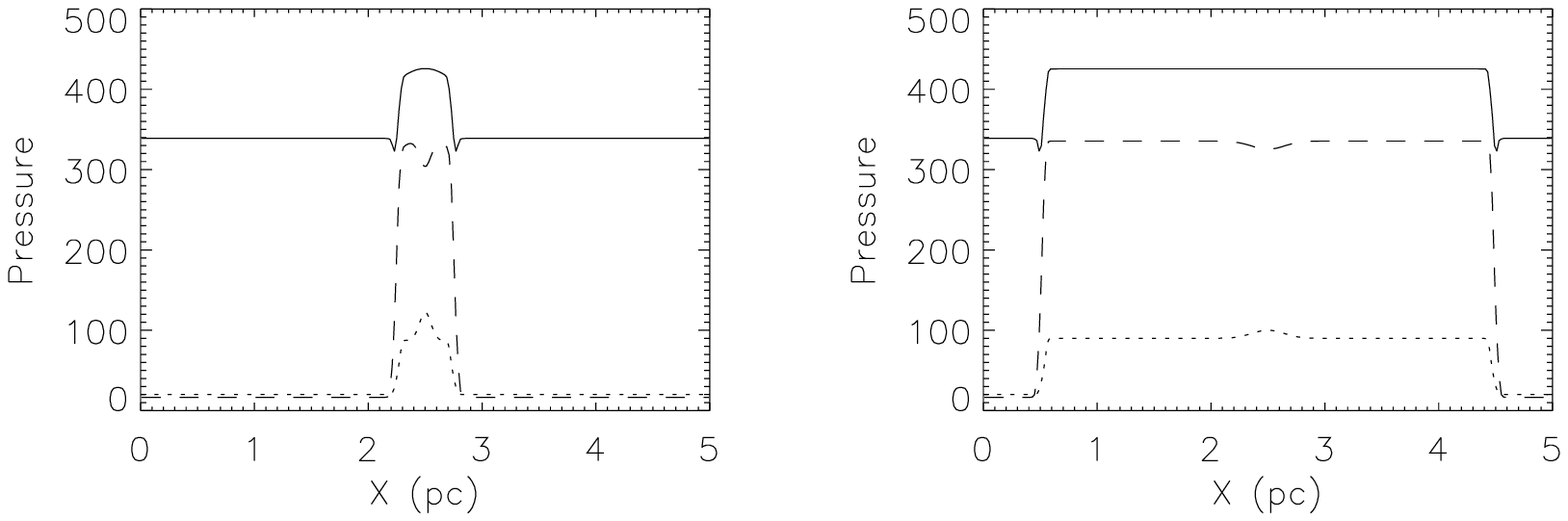}
\caption{$P_\mathrm{tot}$ (solid), $P_B$ (dashed), and $P_\mathrm{gas}$ (dotted) in the colliding clump simulation, corresponding to the earlier (\textit{left}) and later (\textit{right}) stages in Fig.~\ref{CC4}. Note that the magnetic pressure dominates the post-shock region, with the central peak in thermal pressure compensated for by a reduction in the magnetic pressure. Over time, the thermal pressure peak and magnetic pressure valley decline due to diffusion within the post-shock region. In the frame of the (right- or left-ward) expanding shock fronts (not shown), the total pressure in the post-shock region is the same as the total pressure upstream. Units for pressure are $\left[2.3m_\mathrm{H}~\mathrm{cm}^{-3}\right]\left[1~\mathrm{km~s}^{-1}\right]^2$.}
\label{pressure}
\end{figure}

When the two dense clumps meet each other, a strong shock forms (Fig.~\ref{CC3}). Since all fluid variables ($n$, $B_y$, $v_x$) are smooth and continuous prior to shock formation, the features produced are not a consequence of discontinuous initial conditions. This test case eventually evolves to profile similar to that in the simple convergent flow test (Fig.~\ref{CC4}). The central peaks in density and mass-to-flux ratio show up as well. Subsequent evolution leads to a decline in the central peak in $n$ and $n/B_y$ (Fig.~\ref{CC4}).

\subsubsection{Transient C shock Development}

Peaks in density above the ``steady" shock solution have also been observed in other ambipolar diffusion simulations using different MHD codes \citep[e.g.,][]{2009ApJS..181..413C}. In addition, similar transient behavior of C shocks has been noted in models with more complex chemistry implemented \citep[e.g.][]{1998MNRAS.295..672C, 2009MNRAS.395..319V, 2010A&A...511A..41A}. Physically, we believe these peaks arise because the neutrals are effectively ``unmagnetized" when the shock first forms. As a consequence, the neutrals can be very strongly compressed, forming what is seen as a central density peak in Figs.~\ref{cvg1}$-$\ref{CC4}. 

The magnetic field, however, does not follow the initial strong compression of the neutrals. Instead, the overly-compressed neutrals generate higher pressure in the central regions, inhibiting the magnetic flux from getting in. Fig.~\ref{pressure} shows the total pressure $P_\mathrm{tot} = \rho_n {v_n}^2 + P_\mathrm{gas} + P_B$ of the system, where $P_\mathrm{gas} = \rho_n {c_s}^2$ and $P_B =  B^2/8\pi$, with $v_n$ measured in the ``laboratory" frame. These three terms correspond to ${\cal M}^2/r_n$, $r_n$, and ${r_B}^2/\beta$ respectively, in Equation~(\ref{up_down}). Since \textit{Athena} uses the conservative form of the momentum equation ($ \partial \left(\rho v\right) / \partial t + \partial P_\mathrm{tot} / \partial x = 0$), $P_\mathrm{tot}$ must become constant in the post-shock region at late times. For strong shocks, the magnetic pressure term dominates at late times in the post-shock region. At early times, there is a slight depression of the magnetic field strength at the center of the shock, in order to balance the extremely high neutral gas pressure in the density peak. 

Combining the strong neutral compression and slight magnetic exclusion, the mass-to-flux ratio is elevated in the center when a shock forms. The collisions between neutrals and ions will gradually slow down the incoming neutrals and compress ions and magnetic field to the center. Meanwhile, the neutrals in the central peak diffuse outward in order to balance the increasing magnetic pressure and keep the total pressure constant. Eventually, the ions and neutrals interact sufficiently that a steady-state C shock structure develops. The post-shock $n/B_y$ is the same as the upstream value. However, the ambipolar diffusion process takes time, and during the transient stage, a region of very strongly compressed neutrals will be present.

Our finding that there is a transient stage of very strong density compression, with an enhanced ratio of $n/B_y$ or mass-to-magnetic flux, suggests that the very early stage of shock development in GMCs may be particularly important to star formation. The following sections examine this idea further.

\section{Criticality of Clouds}
\label{sec: criticality}

\subsection{Mass-to-flux Ratio}
\label{sec: GammaDef}
The mass-to-flux ratio is a crucial parameter defining whether the magnetic field can support a cloud against its own self-gravity. The critical value of $M/\Phi_B$ for an uniform, spherical cloud has been derived to be $M/\Phi_B \big|_\mathrm{crit}= c_\Phi/\sqrt{G} \approx 0.126/\sqrt{G}$ \citep{1976ApJ...210..326M}. The numerical coefficient $c_\Phi$ differs with the geometry of the cloud: an infinite sheet-like cloud has $c_\Phi = 1/2\pi \approx 0.16$ \citep{1978PASJ...30..671N}, while \cite{1988ApJ...335..239T} found $c_\Phi = 0.17 - 0.18$ for clouds with various $M/\Phi_B$ distributions (see review by \cite{2007ARA&A..45..565M}). Since the value of $c_\Phi$ varies only $\sim 10\%$ with geometry, we choose the commonly-used $c_\Phi = 1/2\pi$ \citep[e.g.][]{2011ApJ...728..123K, 2011MNRAS.414.2511V} as a reference value, while keeping in mind that core geometry is not explicitly defined for our slab system.

Practically, for magnetic field in the $y$-direction the ratio can be written as
\begin{equation}\frac{M}{\Phi} = \frac{\int{\rho dx \cdot L_y L_z}}{\int{B_y dx\cdot L_z}} = \frac{L_y\int{\rho dx}}{\int{B_y dx}} \sim \frac{\Sigma}{\langle B_y\rangle},\end{equation}
where we assume that if a core formed in the post-shock region, its effective length in the $y$-direction, $L_y$, would be comparable to that in the $x$-direction, $L_x$, so that $\langle B_y\rangle = \int{B_ydx}/\int{dx} \sim \int{B_y dx} / L_y$. The mass-to-flux ratio, in units of the critical value $M/\Phi_B\big|_\mathrm{crit} = \left(2\pi\sqrt{G}\right)^{-1}$, is
\begin{equation}
\Gamma\equiv\frac{2\pi\sqrt{G}\cdot\Sigma}{\langle B_y\rangle} = 3.8\left(\frac{N(\mathrm{H})}{10^{21}\mathrm{cm}^{-2}}\right)\left(\frac{\langle B_y\rangle}{\mu\mathrm{G}}\right)^{-1}
\label{GammaDef}
\end{equation}
To convert the column of neutrals in our simulation to $N(\mathrm{H})$, we use $n = n_\mathrm{H_2} + n_\mathrm{He} = 0.6n_\mathrm{H}$. Note that the true value of the normalized mass-to-flux ratio would differ from Equation~(\ref{GammaDef}) by a factor $L_y/L_x$, which could be up to $\sim 2$.

If the mass-to-flux ratio of a prestellar core is larger than the critical value ($\Gamma >1$), i.e., the gravitational force exceeds the magnetic support, the core is supercritical and is eligible for collapse (subject to support by thermal pressure). In contrast, a subcritical core has a mass-to-flux ratio smaller than the critical value ($\Gamma < 1$), and cannot collapse unless it loses magnetic energy in either the strong-gravity mode (the field lines diffuse outward through ambipolar diffusion while gravity holds the gas material together) in which $\Gamma\sim 1$ is required, or the magnetic-dominated mode (neutral mass moves toward the center under the gravitational pull while ambipolar diffusion allows the magnetic field lines to remain stationary) so the mass-to-flux ratio increases.

\subsection{Bonnor-Ebert Sphere}

A typical low-mass prestellar core has $\Sigma_{\mathrm{core}} \sim 1\ \mathrm{M_\odot}/\left[\left(0.1\ \mathrm{pc}\right)^2 \pi\right] \approx 0.007~\mathrm{g\cdot cm^{-2}}$, so that a core with $B_{\mathrm{initial}} \gtrsim 2\pi\sqrt{G}\cdot \Sigma_{\mathrm{core}} \sim 10.7~\mathrm{\mu G}$ may be subcritical.

More precisely, we consider the Bonnor-Ebert sphere radius for a core whose mean density is equal to the post-shock density $\rho_f$,
\begin{equation}
R_\mathrm{BE} = \frac{2.7 c_s}{\left(4\pi G \rho_f\right)^{1/2}}
\label{RBE}
\end{equation}
\citep[e.g.][]{2009ApJ...699..230G}, which is the largest sphere that can be supported by its own internal thermal pressure. 

We note that, from Equation~(\ref{anaL1}), the ratio of the shock thickness to the diameter of a Bonnor-Ebert sphere at the post-shock density is
\begin{equation}
\frac{L_\mathrm{est}}{2R_\mathrm{BE}} \approx \frac{0.7\left(4\pi G \rho_0\right)^{1/2}}{\alpha\rho_{i,0}}\left(\frac{v_0}{c_s}\right)\approx \frac{v_0}{\chi_{i0}c_s}.
\end{equation}
A converging flow bounded by C shocks has breadth at least twice the shock thickness. Under conditions in GMCs where $v_0 /c_s \gtrsim 10$, and $\chi_{i0}\lesssim 10$, this implies that shocks are sufficiently broad that Bonnor-Ebert spheres can fit within the post-shock region. Thus, if magnetic fields are weak enough, cores could grow and collapse in post-shock gas.

The mass-to-flux ratio for a sphere of radius $R_\mathrm{BE}$ in a post-shock magnetized region, without ambipolar diffusion, is
\begin{equation}
\frac{M}{\Phi_B}\bigg|_\mathrm{BE} = \frac{4\pi R^3\rho_f /3}{\pi R^2 B_f} = \frac{4}{3}R_\mathrm{BE}\frac{\rho_f}{B_f},
\end{equation}
with corresponding
\begin{equation}
\Gamma_\mathrm{BE} = \frac{M/\Phi_B\big|_\mathrm{BE}}{M/\Phi_B\big|_\mathrm{crit}} =\frac{8\pi\sqrt{G}}{3}R_\mathrm{BE}\frac{\rho_f}{B_f} = \frac{1.8 c_s}{v_{\mathrm{A},f}} = \frac{1.8 c_s}{r_f^{1/2}v_{\mathrm{A},0}}
\label{GammaBE}
\end{equation}
where
\[v_{\mathrm{A},f}^2 = B_f^2/\left(4\pi\rho_f\right) = r_f v_{\mathrm{A},0}^2\]
using the shock jump conditions. Note that for a strong shock, $r_f\approx\sqrt{2} v_0/v_{\mathrm{A},0}$, so that $\Gamma_\mathrm{BE}\approx 1.5  c_s/\left(v_0 v_{\mathrm{A},0}\right)^{1/2}$, or
\begin{equation}
\Gamma_\mathrm{BE}\approx 0.8\left(\frac{n_0}{100\mathrm{cm^{-3}}}\right)^{0.25}\left(\frac{v_0}{\mathrm{km/s}}\right)^{-0.5}\left(\frac{B_0}{\mu\mathrm{G}}\right)^{-0.5}\left(\frac{T}{10\mathrm{K}}\right)^{-0.5}.
\label{GBEnum}
\end{equation}

If $\Gamma_\mathrm{BE}$ is larger than $1$, a post-shock region of radius $\sim R_\mathrm{BE}$ is dense enough to gravitationally collapse whether or not there is ambipolar diffusion. Otherwise, ambipolar diffusion would be needed for a region of size $\sim R_\mathrm{BE}$ to become supercritical. For the set of shock models we are studying (see Table~\ref{thickness} for inflow parameters), calculated radii and mass-to-flux ratios for Bonnor-Ebert spheres under post-shock conditions without ambipolar diffusion are listed in Table~\ref{Model}. In all cases, $\Gamma_\mathrm{BE}$ is much smaller than $1$, which means no collapse at the BE scale could happen in the post-shock region without significant ambipolar diffusion. More generally, since $\Gamma_\mathrm{BE} \sim c_s / \left(v_0 v_{\mathrm{A},0}\right)^{1/2} \ll 1$ under GMC conditions, most post-shock regions are sufficiently magnetized that gravitational collapse of low-mass cores would be prevented unless ambipolar diffusion occurs. Note that if $\rho > r_f\rho_0$, as is true in the candidate core material for transient C shocks, $R_\mathrm{BE}$ will be lower than the value in the table.

\section{Core Forming Process}
\label{sec: core}

Current theoretical and observational work suggests that shocks produced by supersonic turbulence play a role in compressing gas to form prestellar cores. Our findings that the neutrals are compressed more than the magnetic field during the early stages of shock formation raise an interesting question: Is it possible for a subcritical cloud to form supercritical cores in shocks, which can then gravitationally collapse promptly?

\subsection{Evolution of Overdense Regions}

For a prestellar core to collapse, the region must be dense enough so that self-gravity overcomes the magnetic support. Although self-gravity is not included in the present models, we can make an initial assessment of whether transient C shocks are likely to affect the ability of cores to collapse promptly after they form.

In the context of our converging flow test, we shall define ``candidate core" material to be regions where
\begin{equation}\frac{n}{B_y} > 1.2\times\left( \frac{n}{B_y}\right)_{\mathrm{background}};\end{equation}
the background has uniform $n/B_y$, so the time evolution of this candidate core material is easily calculated. Physically, this candidate core material corresponds to that in the central peak of $n/B_y$ as shown in e.g. Fig.~\ref{cvg1} or Fig.~\ref{CC4}.

For steady shocks with compression factor $r_f$ produced by a two-sided converging flow with inflow speed $v_\mathrm{inflow}$ from both sides, the upstream speed in the shock frame is $v_0 = v_\mathrm{inflow} r_f/ \left(r_f-1\right)$, and $v_\mathrm{shock} = v_\mathrm{inflow}/\left(r_f -1\right)$. The rate at which the column density grows for a steady shock is therefore
\begin{equation}
\frac{\Delta N(\mathrm{H})}{\Delta t} = 2 n_\mathrm{H} v_\mathrm{inflow} = 1.05\times 10^{21} ~\mathrm{cm^{-2} Myr^{-1}} \left(\frac{n_0}{100\mathrm{cm^{-3}}}\right)\left(\frac{v_\mathrm{inflow}}{\mathrm{km/s}}\right),\label{growRate}
\end{equation}
where $n = n_\mathrm{H_2} + n_\mathrm{He} = 0.6n_\mathrm{H}$ is assumed.

\begin{figure}
\epsscale{0.8}
\plotone{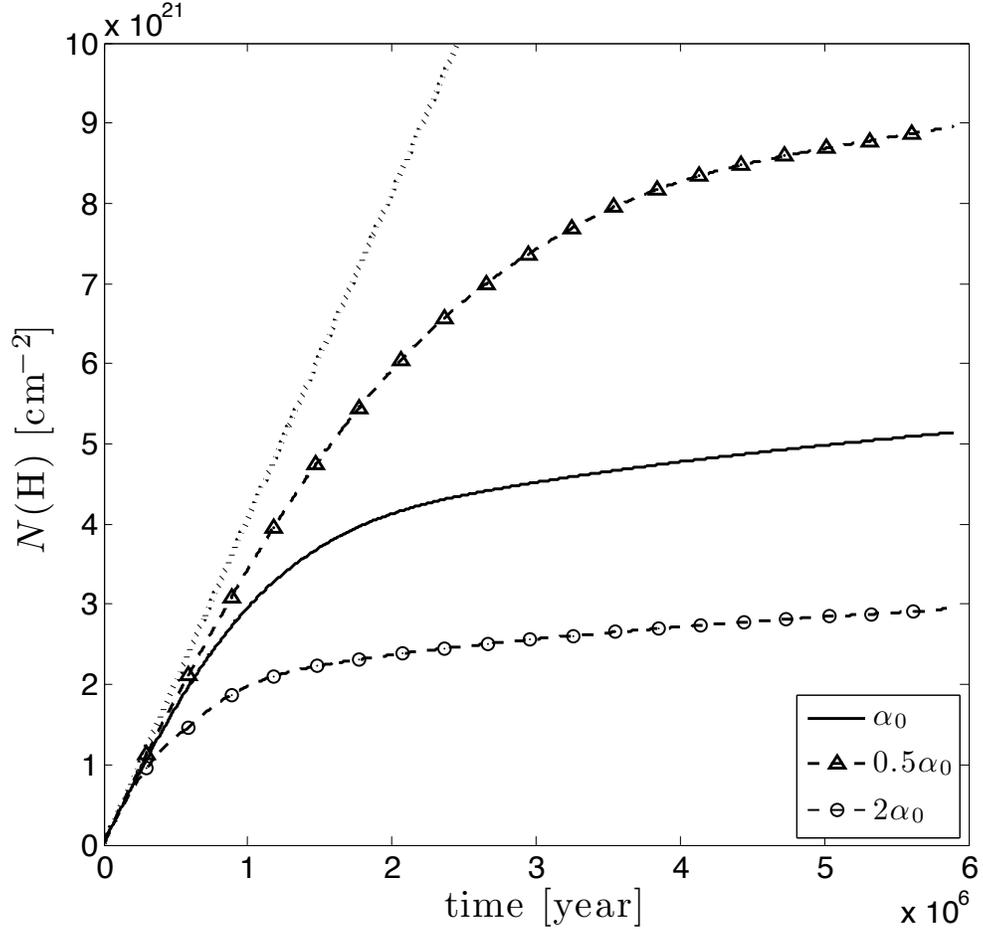}
\caption{Dependence of ``candidate core" material column density on the ion-neutral collisional coefficient, for $\alpha = \alpha_0 = 3.7\times 10^{13}~\mathrm{cm^3 s^{-1} g^{-1}}$, $\alpha = 0.5\alpha_0$, and $\alpha=2\alpha_0$. Also shown (straight dotted line) is the ``kinematic" growth rate $dN(\mathrm{H})/dt = 2n_\mathrm{H}v_\mathrm{inflow}$ for a steady shock. Parameters for this model are $n_0 = 100~\mathrm{cm^{-3}}$, $v_0 = 5~\mathrm{km/s}$, $B_0 = 10~\mathrm{\mu G}$, and $\chi_{i0} = 5$.}
\label{alphaTest}
\end{figure}

For a steady shock, the growth rate of the post-shock column is independent of $\alpha$, the collision coefficient between neutrals and ions. Fig.~\ref{alphaTest} shows evolution of the ``candidate core" column density with different values of $\alpha$, for a model with $n_0 = 100~\mathrm{cm^{-3}}$, $v_\mathrm{inflow} = 3.87~\mathrm{km/s}$, $B_0 = 10~\mu\mathrm{G}$, and $\chi_{i0}=5$. In the very beginning, all of the post-shock material has $n/B_y$ greater than the background value, because the core grows by unimpeded motion of the neutrals which do not ``see" the ions. Thus the column of ``candidate core" material initially follows Equation~(\ref{growRate}), with slope $\Delta N(\mathrm{H})/\Delta t \approx 4\times 10^{21}~\mathrm{cm^{-2}Myr^{-1}}$ (also shown in Fig.~\ref{alphaTest}), independent of $\alpha$.

It is evident that at some point the growth rate of the ``candidate core" column decreases. Physically, the growth rate decreases as ions are pushed into the column by inflowing neutrals, which causes the magnetic flux in the ``candidate core" region to increase more rapidly than the neutral column, and the mass-to-flux ratio to decrease. Therefore, we might expect the growth of the demagnetized column to slow down on a timescale $ t_\mathrm{AD} \sim L_\mathrm{shock}/v_\mathrm{drift}$, the time for neutrals to travel across the shock front under the influence of ions. In timescales short compared to $t_\mathrm{AD}$, neutrals which have arrived at the center were moving fully or partly free from collisions with ions. These neutrals thus contribute to the column with high mass-to-flux ratio. After $t_\mathrm{AD}$, neutrals which have interacted strongly with ions dominate, and the growth rate of the low-magnetization column starts to decrease.

This also corresponds to the timescale for steady C shock structure to develop, and for the fronts surrounding the shocked layer to expand. After this time, neutrals must then travel a greater distance through the condensed magnetic field and ions to stream into the ``candidate core" area. From Fig.~\ref{alphaTest}, the ``saturation" time varies, roughly inversely with the collision coefficient $\alpha$. With Equation~(\ref{thick_ana}) or (\ref{anaL1}) we have $L_\mathrm{shock}$ (and hence $t_\mathrm{AD}$) $\propto 1/\alpha$, consistent with our numerical results.

\subsection{Time Scale and the Mass-to-flux Ratio}
\label{sec:models}

As discussed earlier, our simulation results indicate that during an initial transient period, the neutrals are compressed much more strongly than the magnetic field. Up to a certain time, corresponding to the ambipolar diffusion time scale, the column density of gas with elevated $n/B$ grows. After this time, the profile transitions to that of a steady C shock, with $n/B$ equal to the upstream value.

The ambipolar diffusion time scale should be comparable to the time it takes for neutrals to travel through the thickness of a C-type shock under the influence of ion drag. Therefore we have
\begin{equation} t_\mathrm{AD} \equiv \frac{L_\mathrm{shock}}{\left\langle v_\mathrm{drift}\right\rangle} = \frac{L_\mathrm{shock}}{v_0}\left\langle\left|\frac{1}{r_n} - \frac{1}{r_B}\right|^{-1}\right\rangle, 
\end{equation}
where we have used $v_\mathrm{drift} = \left| v_i - v_n\right| = v_0\left| r_n^{-1} - r_B^{-1}\right|$ (see Equations~(\ref{vn}) and (\ref{vi})). Assuming that $r_n\gg r_B$ over the shock region because a steady-state C shock has not yet formed, and using $\langle r_B\rangle\approx r_f/2$ as an average value, we obtain
\begin{align}
t_\mathrm{AD} &\approx \langle r_B\rangle\frac{L_\mathrm{shock}}{v_0} \approx \frac{r_f}{2}\frac{L_\mathrm{shock}}{v_0} \label{tADapprox}\\
& \approx \frac{2 r_f^{1/2}}{\alpha\rho_{i,0}} \approx \frac{2^{5/4}}{\alpha\rho_{i,0}}\left(\frac{v_0}{v_{\mathrm{A},0}}\right)^{1/2} \label{tADapprox2}.
\end{align}
In the second line, we use the estimate of Equation~(\ref{anaL1}) for the shock thickness, and Equation~(\ref{rfApprox}) for $r_f$ in a strong shock. Note that a similar formula for more general cases with oblique shocks is given in Equation~(\ref{ObtAD}). In dimensional terms, using the analytical approximation Equation~(\ref{thickAna_B2}) to the C shock thickness $L_\mathrm{shock}$, Equation~(\ref{tADapprox}) gives
\begin{align}
t_\mathrm{AD} &\approx 1\times 10^6~\mathrm{yr} \left(\frac{n_0}{\mathrm{100 cm^{-3}}}\right)^{-0.25}\left(\frac{v_0}{\mathrm{km/s}}\right)^{0.5}\left(\frac{B_0}{\mathrm{\mu G}}\right)^{-0.5}\left(\frac{\chi_{i0}}{10}\right)^{-1}\label{tADmodel} \\
& = 0.36\times 10^6~\mathrm{yr} \left(\frac{n_0}{\mathrm{100 cm^{-3}}}\right)^{-0.5}\left(\frac{v_0}{v_{\mathrm{A},0}}\right)^{0.5}\left(\frac{\chi_{i0}}{10}\right)^{-1}.\label{tADmodel2}
\end{align}

The time $t_\mathrm{AD}$ can be compared to the gravitational free fall time
\begin{equation}
t_\mathrm{ff}\left(\rho\right) = \left(\frac{3\pi}{32 G\rho}\right)^{1/2} = 3.4\times 10^6~\mathrm{yr} \left(\frac{n}{100\mathrm{cm^{-1}}}\right)^{-1/2}
\label{tff}
\end{equation}
to give
\begin{equation}
\frac{t_\mathrm{AD}}{t_\mathrm{ff}(\rho)} \approx \left(\frac{\rho}{\rho_0}\right)^{1/2}\left(\frac{v_0}{v_{\mathrm{A},0}}\right)^{1/2} {\chi_{i0}}^{-1}.
\end{equation}
The post-shock gas has density $\rho_f = r_f\rho_0$ with $r_f \approx \sqrt{2}\left(v_0/v_{\mathrm{A},0}\right)$ (see Equation~(\ref{rfApprox})), which means that $t_\mathrm{AD}/t_\mathrm{ff}(\rho_f)\sim \left(v_0 / v_{\mathrm{A},0}\right)\chi_{i0}^{-1}$. During the transient stage, $\rho_t>\rho_f$, so $t_\mathrm{ff}\left(\rho_t\right)<t_\mathrm{ff}\left(\rho_f\right)$, implying $t_\mathrm{AD}/t_\mathrm{ff}\left(\rho_t\right) > \left(v_0 / v_{\mathrm{A},0}\right)\chi_{i0}^{-1}$. Thus, for strong shocks ($v_0 / v_{\mathrm{A},0} \gtrsim 10$) and low ionization conditions ($\chi_{i0}\lesssim 10$), the transient duration $t_\mathrm{AD}$ will exceed the time $t_\mathrm{ff}\left(\rho_t\right)$ for post-shock perturbations to develop into collapsing cores.

If the growth rate of the neutral column density is $dN_\mathrm{H}/dt \approx 2 n_\mathrm{H} v_0$ (see Equation~(\ref{growRate})), the final (maximum) value of the mass-to-flux ratio should be
\begin{equation}
\frac{M}{\Phi_B}\bigg|_\mathrm{final} \approx \frac{1.4 m_\mathrm{H} dN_\mathrm{H}/dt \times t_\mathrm{AD}}{B_\mathrm{final}} \approx \frac{2 \rho_0 v_0 \times L_\mathrm{shock}r_f/ \left(2 v_0\right)}{r_f B_0} = \frac{\rho_0 L_\mathrm{shock}}{B_0}.
\label{MdBfinal}
\end{equation}
This estimate of the mass-to-flux ratio inside the pre-collapsing core depends only on the upstream density and magnetic field, and the steady-state C shock thickness. This can be evaluated (using Equation~(\ref{thickAna_B2})) to give $\Gamma_\mathrm{final} \equiv 2\pi\sqrt{G}\left(M/\Phi_B\right)_\mathrm{final}$:
\begin{align}
\Gamma_\mathrm{final} &\approx 0.41 \left(\frac{n_0}{\mathrm{100 cm^{-3}}}\right)^{0.25}\left(\frac{v_0}{\mathrm{km/s}}\right)^{0.5}\left(\frac{B_0}{\mathrm{\mu G}}\right)^{-0.5}\left(\frac{\chi_{i0}}{10}\right)^{-1}\label{GADmodel} \\
& = 0.14 \left(\frac{v_0}{v_{\mathrm{A},0}}\right)^{0.5}\left(\frac{\chi_{i0}}{10}\right)^{-1}.\label{GADmodel2}
\end{align}
Note that true $\Gamma$ would differ by a factor $L_y/L_x$ from Equations~(\ref{GADmodel}) and (\ref{GADmodel2}); i.e.~up to factor $\sim 2$ larger.

Combining Equations~(\ref{tADapprox2}), (\ref{tff}), and (\ref{MdBfinal}), we have
\begin{equation}
t_\mathrm{AD} = \left(\frac{32}{6\pi^2}\right)^{1/2}\Gamma_\mathrm{final} t_\mathrm{ff}\left(\rho_0\right) = 0.74~\Gamma_\mathrm{final}t_\mathrm{ff}\left(\rho_0\right).
\label{tADGfinal}
\end{equation}
Thus, shocks that are able to reach $\Gamma_\mathrm{final}\sim 1$ through transient ambipolar diffusion will do so on a timescale comparable to the gravitational time $t_\mathrm{ff}\left(\rho_0\right)$ of the large-scale cloud. Since the large-scale dynamical timescale ($\sim t_\mathrm{ff}(\rho_0)$ for a self-gravitating cloud) determines the correlation time of the flows that create shocks, this means that shocks will be sustained long enough for diffusion to occur. 

If $v_0 \gtrsim v_{\mathrm{A},0}$ and $\chi_{i0}\sim 1$, from Equation~(\ref{GADmodel2}) the candidate core will have $\Gamma_\mathrm{final}$ exceeding unity. In this situation, a core would be able to collapse promptly, without an extended period of ambipolar diffusion, since $t_\mathrm{AD}/t_\mathrm{ff}\left(\rho_t\right) \gtrsim \left(v_0/v_{\mathrm{A},0}\right)^{1/2}\Gamma_\mathrm{final} > \Gamma_\mathrm{final}$. In GMCs, the ionization fraction is dependent on chemical processes, with $\chi_{i0}\sim 1-20$ \citep{2010ApJ...720.1612M}. The turbulent flow speed will not exceed $\sim 10$~km/s under realistic conditions, and $n_0 \sim 10^2-10^3$~cm$^{-3}$, typically\footnote{Keep in mind that some combinations of parameters are not astronomically realistic; e.g.~high $v_0$ is unlikely to have low $\chi_{i0}$, and high $n_0$ is unlikely to have low $B_0$.}. Therefore, $\Gamma_\mathrm{final}$ will exceed $1$ (see Equation~(\ref{GADmodel})) only if the upstream magnetic field in the direction parallel to the shock front is moderate, probably $\lesssim 10~\mu\mathrm{G}$. The line-of-sight magnetic field strengths in molecular clouds with density $\lesssim 10^3~\mathrm{cm}^{-3}$, however, can vary over $\sim 5 - 25~\mu\mathrm{G}$ \citep{1999ApJ...520..706C, 2010ApJ...725..466C}. If the total magnetic field strength (which is always $\geq B_\mathrm{LOS}$) exceeds $\sim 20\mu\mathrm{G}$, then in order to reach $\Gamma_\mathrm{final}$ close to $1$ so that pre-collapse cores can develop efficiently, converging flows with $\mathbf{v}_0$ aligned $\lesssim 30^\circ$ to $\mathbf{B}_\mathrm{cloud}$ are favored. 

In addition, we note that for $\sigma_\mathrm{3D,~cloud}$ and $v_{\mathrm{A},~\mathrm{cloud}}$ the 3D turbulent velocity dispersion and mean-field Alfv\'{e}n speed in a cloud, a gravitationally-bound (or virialized) cloud has
\begin{equation}
\Gamma_\mathrm{cloud} \sim \frac{\sigma_\mathrm{3D,~cloud}}{v_{\mathrm{A},~\mathrm{cloud}}},
\end{equation}
so that
\begin{equation}
\Gamma_\mathrm{final} \sim \frac{1}{\chi_{i0}}\left(\frac{v_0}{\sigma_\mathrm{3D,~cloud}}\right)^{1/2}\left(\frac{B_\mathrm{cloud}}{B_0}\right)^{1/2}\Gamma_\mathrm{cloud}^{1/2}.
\label{GADcloud}
\end{equation}
The strongest shocks will have $v_0\sim\sigma_\mathrm{3D,~cloud}$. These regions will be able to reach $\Gamma_\mathrm{final} \sim 1$ if the cloud is sufficiently supercritical ($\Gamma_\mathrm{cloud} \gg 1$), the ionization fraction is sufficiently low ($\chi_{i0} \sim 1$), and/or the magnetic field parallel to the shock front is weaker than the mean field threading the cloud ($B_\mathrm{cloud} / B_0 > 1$). Again, with realistic $\chi_{i0}$ and $\Gamma_\mathrm{cloud}$, the most favorable circumstance for ambipolar diffusion to yield $\Gamma_\mathrm{final} > 1$ is if the inflow $\mathbf{v}_0$ is aligned locally towards $\mathbf{B}_\mathrm{cloud}$ so that $B_\mathrm{cloud}/B_0 >1$.\footnote{We have investigated oblique shocks with nonzero $B_{x,0} = B_\mathrm{cloud}\cos\theta = B_0 \cot \theta$ in Appendix~\ref{sec:appendix}, where $B_0 = B_{y,0} = B_\mathrm{cloud}\sin\theta$ is the magnetic component parallel to the shock front. Equation~(\ref{ObGamma}) gives an approximation of $\Gamma_\mathrm{final}$ as a function of $\theta$ (the angle between $\mathbf{B}_\mathrm{cloud}$ and $\mathbf{v}_0$). Since there is no strong dependence of $\Gamma_\mathrm{final}$ on $\theta$, our 1-D results (Equations~(\ref{GADmodel}), (\ref{GADmodel2}), and (\ref{GADcloud})) are applicable in most cases with nonzero $B_{x,0}$. }

Even if post-shock regions are subcritical, transient ambipolar diffusion significantly increases the mass-to-flux ratio compared to the value that would hold in ideal MHD. A measure of the importance of this effect is the ratio between $\Gamma_\mathrm{final}$ and $\Gamma_\mathrm{BE}$ in the post-shock region. From Equations~(\ref{GADmodel}) and (\ref{GBEnum}),
\begin{equation}
\frac{\Gamma_\mathrm{final}}{\Gamma_\mathrm{BE}} \sim 5 \left(\frac{v_0}{\mathrm{km/s}}\right)\left(\frac{T}{10\mathrm{K}}\right)^{-1/2}\chi_{i0}^{-1} \sim \frac{{\cal M}}{\chi_{i0}}
\label{GfGBE}
\end{equation}
is predicted. The turbulent motions in clouds can achieve ${\cal M}\sim 50$. With $\chi_{i0}\sim 1-20$, a significant enhancement in the mass-to-flux ratio can be expected due to transient ambipolar diffusion.

\subsection{Simulation Results}

\begin{deluxetable}{cccccccccccccr}
\tablecolumns{14}
\tabletypesize{\footnotesize}
\tablecaption{Results for Transient Mass-to-flux Enhancement \label{Model}}
\tablewidth{0pt}
\tablehead{
\colhead{} & \colhead{} &
\multicolumn{2}{c}{$t_\mathrm{AD}$} &  \colhead{}  &
\multicolumn{2}{c}{$\Gamma_\mathrm{final}$}  & \colhead{}&
\colhead{$L_\mathrm{core}$} & \colhead{} &
\multicolumn{2}{c}{BE Sphere~\tablenotemark{a}} &  \colhead{}  &
\colhead{}\\

\cline{3-4} \cline{6-7} \cline{11-12}
\colhead{} & \colhead{} &
\colhead{eq.~(\ref{tADmodel})} & \colhead{result~\tablenotemark{b}} &\colhead{} & 
\colhead{} & \colhead{} & \colhead{} &
\colhead{at $2t_\mathrm{AD}$~\tablenotemark{c}} & \colhead{} &
\colhead{$R_\mathrm{BE}$} &\colhead{$\Gamma_\mathrm{BE}$} & \colhead{} &
\colhead{$\Gamma_\mathrm{2t_\mathrm{AD}}/$}\\

\colhead{Model} &  \colhead{} &
\multicolumn{2}{c}{($10^6$ years)}& \colhead{}  &
\colhead{eq.~(\ref{GADmodel})} & \colhead{$\Gamma_\mathrm{2t_\mathrm{AD}}$} & \colhead{}&
\colhead{(pc)} & \colhead{} &
\colhead{(pc)} & \colhead{}  & \colhead{} &
\colhead{$\Gamma_\mathrm{BE}$}
}
\startdata
N01 &&1.42&  1.12 && 0.55 & 0.37 && 0.58 &&0.46 & 0.12 &&  3.08\\
N03 &&1.10&  1.11 && 0.74 & 0.64 && 0.34 &&0.20 & 0.15 &&  4.27\\
N05 &&0.98&  1.09 && 0.85 & 0.78 && 0.27 &&0.13 & 0.17 &&  4.59\\
N08 &&0.88&  1.07 && 0.96 & 0.95 && 0.23 &&0.09 & 0.19 &&  5.00\\
N10 &&0.83&  0.90 && 1.02 & 1.07 && 0.21 &&0.08 & 0.20 &&  5.35\\
V04 &&1.07&  0.94 && 0.59 & 0.44 && 0.40 &&0.31 & 0.16 &&  2.75\\
V06 &&1.33&  1.42 && 0.73 & 0.60 && 0.42 &&0.25 & 0.13 &&  4.62\\
V08 &&1.56&  1.86 && 0.86 & 0.76 && 0.42 &&0.21 & 0.11 &&  6.91\\
V10 &&1.75&  1.88 && 0.96 & 0.91 && 0.43 &&0.19 & 0.10 && 9.10\\
V12 &&1.94&  2.13 && 1.06 & 1.06 && 0.43 &&0.17 & 0.09 && 11.78\\
B02 && 2.88& 4.93 && 1.58 & 2.83 && 0.51 && 0.12 & 0.31 && 9.13\\
B04 && 1.98& 2.64 && 1.09 & 1.35 && 0.44 && 0.17 & 0.22 && 6.14\\
B06 &&1.59&  2.15 && 0.87 & 0.89 && 0.43 && 0.21 & 0.18 && 4.94\\
B08 &&1.36&  1.58 && 0.75 & 0.66 && 0.42 && 0.24 & 0.15 && 4.40\\
B10 &&1.21&  1.08 && 0.66& 0.52 & & 0.42 && 0.27 & 0.14 && 3.71\\
B12 &&1.09&  0.90 && 0.60 & 0.43 && 0.42 && 0.30 & 0.13 && 3.31\\
B14 &&1.01&  0.79 && 0.55 & 0.37 && 0.40 && 0.33 & 0.12 && 3.08\\
X01 &&6.03&  5.78 && 3.32 & 2.63 && 2.03 &&0.27 &0.14 &&  18.79\\
X06 &&1.01&  0.89 && 0.55 & 0.43 && 0.34 &&0.27 & 0.14 &&  3.07\\
X10 &&0.60&  0.57 && 0.33 & 0.25 && 0.20 &&0.27 & 0.14 && 2.78\\
X15 &&0.40&  0.48 && 0.22 & 0.16 && 0.14 &&0.27 & 0.14 && 1.79\\
X20 &&0.30&  0.45 && 0.17 & 0.12 && 0.12 &&0.27 & 0.14 && 0.86\\
\enddata

\tablenotetext{a}{Computed for post-shock conditions without ambipolar diffusion (see Equations~(\ref{RBE}) and (\ref{GammaBE})).}
\tablenotetext{b}{Defined as when the slope of the $\Gamma$~vs.~time curve drops to $25\%$ of its maximum value.}
\tablenotetext{c}{$L_\mathrm{core}\equiv N/\langle n\rangle$ in ``candidate core" region with enhanced $n/B$.}
\end{deluxetable}

\begin{figure}
\epsscale{0.8}
\plotone{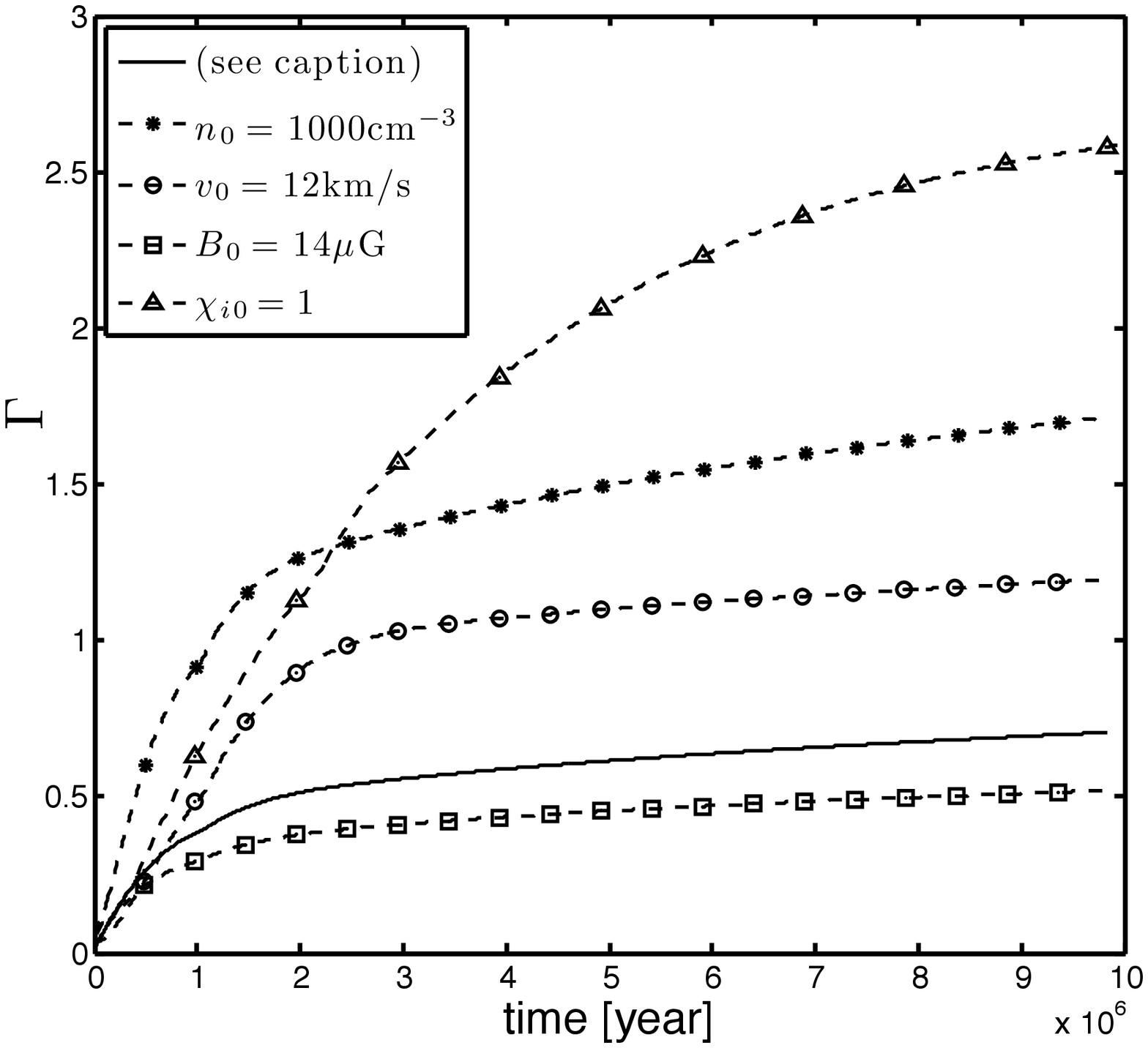}
\caption{Time evolution of the normalized central mass-to-flux ratio $\Gamma$ in the shocked gas. The parameters are $n_0 = 200~\mathrm{cm^{-3}}$, $v_0 = 5~\mathrm{km/s}$, $B_0 = 10~\mu\mathrm{G}$, and $\chi_{i0} = 5$ (solid line), with modifications as noted in the key.}
\label{spcTest}
\end{figure}

The estimates of Equations~(\ref{tADmodel}) and (\ref{GADmodel}) can be compared to the ambipolar diffusion time and mass-to-flux ratio as measured directly from time-dependent numerical simulations. Examples showing evolution of the measured $\Gamma$ for several different parameter values are shown in Fig.~\ref{spcTest}.

To read the ambipolar diffusion time scale from simulations, recall that the growth rate of the mass-to-flux ratio inside the core decreases at time $\sim t_\mathrm{AD}$. We adopt a definition of $t_\mathrm{AD}$ as the time when the slope of the $\Gamma$~vs.~time curve drops to $25\%$ of its maximum value. For each simulation, we measure the mass-to-flux ratio $\Sigma/\langle B_y\rangle$ at time $t = 2t_\mathrm{AD}$, and define this mass-to-flux ratio inside the central peak (multiplied by $2\pi\sqrt{G}$) as $\Gamma_\mathrm{final}$.

Table~\ref{Model} shows the predicted values of $t_\mathrm{AD}$ and $\Gamma_\mathrm{final}$ from Section~\ref{sec:models}, as well as the simulation results for these quantities. The measured ambipolar diffusion time scale is $\sim 0.3-3$~Myr. Our model predicts the ambipolar diffusion time scale very well: the RMS value of $\left(t_\mathrm{AD,~pred} - t_\mathrm{AD,~sim}\right)/ t_\mathrm{AD,~sim}$ is $0.19$, and the range of $\left(t_\mathrm{AD,~pred} - t_\mathrm{AD,~sim}\right)/ t_\mathrm{AD,~sim}$ is $-0.42$ to $0.28$. The measured mass-to-flux ratios deviate from predicted values somewhat more, with a range of $\left(\Gamma_\mathrm{final,~pred} - \Gamma_\mathrm{final,~sim}\right)/ \Gamma_\mathrm{final,~sim}$ $-0.44$ to $0.49$, and RMS value $0.28$. The typical size $L_\mathrm{core}\equiv N/\langle n\rangle$ of the region with enhanced mass-to-flux ratio at time $2 t_\mathrm{AD}$ is $\sim 0.2-0.6$~pc (Table~\ref{Model}, column 6).

In most of our simulations, the mass-to-flux ratios are higher than $0.6$ (see column 5 of Table~\ref{Model}), with some cases (N10, V12, B02, B04, and X01) reaching $\Gamma_{2t_\mathrm{AD}} > 1$. Recall that we assumed the effective length of the system is comparable in all directions when we define $\Gamma$ in the candidate core material (see Section~\ref{sec: GammaDef}). This means that the real mass-to-flux ratio would differ from our measured $\Gamma_{2t_\mathrm{AD}}$ by a factor $L_y/L_x$. Since cores may have axis ratios $\sim 2:1$, the measured $\Gamma_{2t_\mathrm{AD}}$ may be underestimated by a factor up to $\sim 2$. Therefore the fact that almost all models have $\Gamma_{2t_\mathrm{AD}}$ close to $1$ shows that C shock transients may lead to supercritical cores quite frequently.

The models with $\Gamma_{2t_\mathrm{AD}} > 1$ confirm our prediction that small values of $B_0$ are crucial for forming supercritical cores (otherwise uncommonly high neutral density/inflow speed or extremely low ionization fraction may become necessary). Given the limits on physical conditions in clumps within GMCs (see discussion in Section~\ref{sec:models}), prompt supercritical core formation would preferentially occur if the inflow direction is aligned relatively close to the magnetic field. A study of oblique shocks using similar analysis to that in the previous section is performed in Appendix~\ref{sec:appendix}, where we show that the transient behavior of C shocks is insensitive to the component of magnetic field parallel to inflow velocity, so that our 1-D model is qualitatively applicable in cases with more general geometry. 

We also list the value of $\Gamma_{2t_\mathrm{AD}}/\Gamma_\mathrm{BE}$ in Table~\ref{Model} for all our numerical models. In most cases, this ratio is greater than $2$, and the average value is $\sim 5.4$. This means that for essentially all reasonable parameters, transient ambipolar diffusion will be important in enhancing the mass-to-flux ratio for forming cores. Since $\Gamma_{2t_\mathrm{AD}}$ is close to $1$ in many situations, and $\Gamma_{2t_\mathrm{AD}}/\Gamma_\mathrm{BE}$ is large, transient ambipolar diffusion during core formation clearly plays an important role in setting the stage for subsequent core evolution.

\section{Summary}
\label{sec: summary}

Ambipolar diffusion is an important phenomenon in interstellar clouds, which are strongly magnetized in the sense $v_\mathrm{A}\sim v_0\gg c_s$, but are poorly ionized. Supersonic turbulence creates shocks, but ambipolar diffusion between ions and neutrals spreads these shocks out. The thickness of C-type shocks depends on the inflow velocity, density, the magnetic field strength, and the ionization fraction. Although C shocks are normally studied in the steady-state limit, their early transient development is quite interesting. During this transient stage, the central compression of neutrals is strongly enhanced because they are effectively ``unmagnetized." The time and space scales of these transients make them important to the structure and dynamics within GMCs. The transient duration is comparable to the drift time across the C shock thickness ($\sim 0.1-1$~pc), typically $\sim 0.1-1~\mathrm{Myr}$ for GMC conditions.

For star formation, ambipolar diffusion is usually analyzed in the context of slow evolution leading to gravitational collapse in magnetically-supported clouds. However, our results show that since neutrals can stream through field lines in shocks because of ambipolar diffusion, magnetically-supercritical cores may form due to C shock transients. During the transient, strong central compression of neutrals enhances $n/B$ compared to steady-state values. If the compression and duration of the transient are sufficient, the central post-shock region may become supercritical and collapse gravitationally to make a prestellar core before it re-expands.

For both the traditional picture of supercritical core formation and the scenario we propose, the magnetic field remains relatively stationary while the neutrals move inward, within the high density regions. For the traditional picture, the inward neutral motions are due to small-scale self-gravity within the core. For shock-induced core formation, the inward motions of neutrals owes to large-scale converging supersonic flows within GMCs (which may ultimately be driven by large-scale self-gravity within the cloud).

Transient ambipolar diffusion is particularly important because without it, post-shock regions in GMCs typically have very small mass-to-magnetic flux ratios. Thus, the regions with the shortest gravitational timescales (at high density, due to shocks) would be prevented from collapsing by magnetic fields, which are also enhanced by shocks. Our numerical simulations show a peak in the mass-to-flux ratio, produced by transient ambipolar diffusion. For strong shocks ($v_0/v_{\mathrm{A},0}$ sufficient) and low enough ionization fraction, our results suggest that supercritical cores can be produced.

Based on our simulation results and analyses, our main conclusions are as follows: 
\begin{enumerate}

\item
The dominant factors determining the ionization fraction in molecular clouds are ionizing cosmic rays and gas-phase recombination. We derive steady-state equations for C shocks including ionization and recombination (Equations~(\ref{drndx}) and (\ref{dxidx})). Analyzing the solutions of these equations (Fig.~\ref{compStruc}), we find that ionization-recombination equilibrium is generally an excellent approximation, and for this regime is much better than the widely-applied frozen-in condition. For equilibrium ionization, $\rho_i \propto \rho_n^{1/2}$ \citep{1976PASJ...28..355N, 1979PASJ...31..697N} so that $r_i = {r_n}^{1/2}$ in our notation, and Equation~(\ref{govEq}) governs steady C shocks.

\item 
We have solved the steady C shock ODE over a parameter range of upstream neutral density $n_0 = 10^2-10^3$~cm$^{-3}$, inflow speed $v_0 = 2-15$~km/s, upstream magnetic field strength $B_0 = 2-15$~$\mu\mathrm{G}$, and ionization parameter $\chi_{i0} = 1-21$ ($\chi_{i0}$ is defined in Equations~(\ref{rec-ion})$-$(\ref{chiDef})). Using a multilinear fit, we obtain an expression for the C shock thickness (Equation~(\ref{thick_num})), in terms of these parameters. We also obtain an analytic expression for the C shock thickness (Equation~(\ref{thickAna_B1})), which is in excellent agreement with the numerical result. The dependence $L_\mathrm{shock}\propto \left(v_0 v_{\mathrm{A},0}\right)^{1/2} / \left(\alpha\rho_{i,0}\right)$ can be understood based on the requirement for momentum transfer mediated by ion-neutral collisions. Our result for the C shock thickness is comparable to previous estimates \citep[e.g.][]{1993ARA&A..31..373D}, although the parameter dependence differs from the case of ``frozen-in" ions \citep{1990MNRAS.246...98W, 2006ApJ...653.1280L}.

\item
During the transient stage of C shocks, the central column density of gas with enhanced mass-to-flux ratio initially grows kinematically (Equation~(\ref{growRate})), but this slows after a time comparable to the ion-neutral drift time $t_\mathrm{AD} \approx L_\mathrm{shock} / v_\mathrm{drift}$ across the C shock (Fig.~\ref{alphaTest} and Equation~(\ref{tADapprox})). The duration of the transient from our numerical models (see Table~\ref{Model}) is similar to our analytic estimate, $0.1-1~\mathrm{Myr}$ for the regime we have studied (see Equations~(\ref{tADmodel})$-$(\ref{tADmodel2})). Although the present models do not include self-gravity, the duration of the transient C shock is comparable to the time needed for prestellar cores to collapse, from both observations \citep[e.g.][]{2007prpl.conf...33W, 2009ApJS..181..321E} and numerical simulations \citep[e.g.][]{2011ApJ...729..120G}.

\item
Our finding of rapid initial enhancement in density and mass-to-flux ratio is consistent with the results of \cite{2008ApJ...679L..97K}, for somewhat different parameter regime. Their simulations included self-gravity, and they also pointed out that with appropriate parameters, collapsing cores may form due to the initial compression. Our work helps to explain the physics behind the rapid collapse they identified, and more generally provides insight into other numerical studies of turbulence-accelerated, magnetically-regulated star formation \citep[e.g.][]{2004ApJ...609L..83L}.

\item
Over the transient time $t_\mathrm{AD}$, a column $\sim 2n_0 v_0 t_\mathrm{AD}$ of ``candidate core" material accumulates. By taking the ratio with the post-shock magnetic field strength, we can estimate the mass-to-flux ratio of this dense material. Equation~(\ref{GADmodel}) gives an estimate of the final dimensionless mass-to-flux estimate, $\Gamma$, which is similar to numerical measures of $\Gamma_\mathrm{final}$ (Table~\ref{Model}). The relatively high mass-to-flux ratios we find may explain the weak magnetic fields observed in dense cores \citep{2008ApJ...680..457T}. Without ambipolar diffusion, the post-shock mass-to-flux ratio on the scale of a Bonnor-Ebert sphere ($\Gamma_\mathrm{BE} \approx 1.5c_s / \left(v_0v_{\mathrm{A},0}\right)^{1/2}$) would be much smaller than the critical value. In contrast, the mass-to-flux ratio in the candidate core material produced within transient C shocks is several times larger than $\Gamma_\mathrm{BE}$ (Table~\ref{Model} and Equation~(\ref{GfGBE})). This large enhancement shows the significance of ambipolar diffusion during shock-induced core formation.

\item
In transient shocks that can produce $\Gamma_\mathrm{final}\gtrsim 1$, magnetically supercritical cores can form and collapse rapidly. Shocks that can reach $\Gamma_\mathrm{final}\gtrsim 1$ have $t_\mathrm{AD}$ comparable to the gravitational free-fall time of the larger-scale cloud (Equation~(\ref{tADGfinal})). Thus, shock-induced ambipolar diffusion is rapid, wherever it occurs.

\item
The most favorable conditions for forming gravitationally bound cores in cold, turbulent, magnetized clouds are strong shocks ($v_0 \gg v_{\mathrm{A},0}$) in regions with low ionization fraction ($\chi_{i0}\sim 1$). Equation~(\ref{ObGamma}) in the Appendix gives the final mass-to-flux ratio for the case of oblique shocks; the result is similar to that with the same component of the magnetic field parallel to the shock front (Equation~(\ref{GADmodel})). Considering realistic conditions in molecular clouds, converging flows with $\mathbf{v}_\mathrm{inflow}\perp\mathbf{B}_\mathrm{cloud}$ will have relatively low post-shock density and post-shock mass-to-flux ratio, and the post-shock gas layers formed will be unfavorable for star formation. Cases where $\mathbf{v}_\mathrm{inflow}$ and $\mathbf{B}_\mathrm{cloud}$ are more aligned are more favorable for reaching $\Gamma_\mathrm{final}>1$ (Equation~(\ref{GADcloud})). Further observations of the directions of magnetic fields relative to observed gas filaments with or without embedded cores will test whether these orientation effects are indeed important. If orientation of shocks is in fact important in producing cores that can collapse, this may help explain the observed inefficiency of star formation in GMCs.

\end{enumerate}

While the present models are extremely useful for explaining the phenomenon of transient ambipolar diffusion, simulations of more generalized cases are required to support the scenario of prompt supercritical core formation in shocks. Three-dimensional simulations of systems with oblique shocks, including self-gravity of the gas, would be immediately helpful. In addition, a more realistic core-forming environment can be examined by adding nonlinear turbulence to the inflow velocity field. Further simulations along these lines, together with observations  probing density and magnetic structure in filaments and cores at different stages, will improve understanding of what precipitates star formation.

\acknowledgements
We are grateful to the referee for a very thorough and helpful report. This work was supported by NASA under grant NNX10AF60G.

\appendix
\section{Oblique C shocks}
\label{sec:appendix}

The main text considers a 1-D system with velocities and magnetic fields perpendicular to each other, for simplicity. We expect that our results will qualitatively hold for more general geometry. Here, we show that under certain conditions, our results can quantitatively be applied to oblique C-type shocks. 

In the following, we shall consider a plane-parallel shock in the standard shock frame, using the same coordinate system as before. The shock front is in the $y$-$z$ plane, the upstream flow is along the $x$-direction ($\mathbf{v}_{0} = v_{0}\hat{\mathbf{x}}$), and the upstream magnetic field is now in the $x$-$y$ plane ($\mathbf{B}_\mathrm{cloud} = B_{x,0} \hat{\mathbf{x}} + B_{y,0} \hat{\mathbf{y}}$), at an angle $\theta$ to the inflow ($B_{y,0}/B_{x,0} = \tan\theta $).

For steady, plane-parallel shocks, $\partial_t = \partial_y = \partial_z = 0$. From the mass and momentum conservation equations for neutrals (Equations~(\ref{mCon})$-$(\ref{NeuMom})), we have
\begin{subequations}
\begin{align}
\frac{d}{dx}\left(\rho_n v_{n,x}\right) &= 0,\\
\frac{d}{dx} \left(\rho_n v_{n,x}^2 + c_{s}^2 \rho_n\right) &= \alpha\rho_i\rho_n\left(v_{i,x} - v_{n,x}\right),\\
\frac{d}{dx}\left(\rho_n v_{n,x}v_{n,y}\right) &= \alpha\rho_i\rho_n\left(v_{i,y} - v_{n,y}\right).\label{ObNcross}
\end{align}
\end{subequations}
Similarly, the momentum equation for ions and the magnetic induction equation (Equations~(\ref{IonMom})$-$(\ref{induc})) are (with the strong coupling approximation)
\begin{subequations}
\begin{align}
\frac{1}{8\pi}\frac{d}{dx} B_y^2 &=\alpha\rho_i\rho_n\left(v_{n,x}-v_{i,x}\right),\\
\frac{B_x}{4\pi}\frac{d}{dx}B_y &= -\alpha\rho_i\rho_n\left(v_{n,y}-v_{i,y}\right),\label{ObBcross}\\
v_{i,x}B_y - v_{i,y}B_x &= const. = v_0B_{y,0}.
\end{align}
\end{subequations}
Note that $B_x = const. = B_{x,0}$ in plane-parallel shocks, since $\nabla\cdot\mathbf{B} = 0$.

By defining our parameters as
\begin{equation}
r_n \equiv \frac{\rho_n}{\rho_{0}} = \frac{v_0}{v_{n,x}},\ \ \ r_{ix} \equiv  \frac{v_0}{v_{i,x}},\ \ \ r_B \equiv r_{B_y}= \frac{B_y}{B_{y,0}},
\end{equation}
and
\begin{equation}
{\cal M} \equiv {\cal M}_x = \frac{v_{0}}{c_{s}},\ \ \  \frac{1}{\beta_y} \equiv \frac{B_{y, 0}^2}{8\pi\rho_{0}c_{s}^2} = \frac{1}{2}\left(\frac{v_{\mathrm{A}y,0}}{c_{s}}\right)^2,
\end{equation}
and applying the ionization-recombination equilibrium ($\rho_i \propto \rho_n^{1/2}$), Equations~(\ref{drB2dx}) and (\ref{drn+dB2}) become
\begin{align}
\frac{d}{dx} r_B^2 &= -\beta_y\frac{\alpha\rho_{i,0}}{v_0}{\cal M}^2 r_n^{3/2}\left(\frac{1}{r_{ix}}-\frac{1}{r_n}\right)\\
\frac{d}{dx}\left(\frac{{\cal M}^2}{r_n}\right) &+ \frac{d}{dx}\left(r_n\right) = -\frac{1}{\beta_y}\frac{d}{dx} r_B^2. \label{ObdrbdB2}
\end{align}
Since Equation~(\ref{ObdrbdB2}) is the same as Equation~(\ref{drn+dB2}) with $\beta\rightarrow\beta_y$, the $r_B$~vs.~$r_n$ relation for oblique shocks is the same as in shocks with $B_x=0$ (Equation~(\ref{rB})). In addition, in the post-shock regime, $r_{n,f} = r_{ix,f}$. From Equations~(\ref{ObNcross}) and (\ref{ObBcross}), the neutral velocity parallel to the front is given by 
\begin{equation}
\frac{v_{n,y}}{v_0} = \frac{1}{\tan\theta}\left(\frac{v_{\mathrm{A}y,0}}{v_0}\right)^2\left(r_B-1\right).
\end{equation}
The governing equation in the direction perpendicular to the shock front now becomes
\begin{equation}
\frac{d}{dx} r_n = \frac{-D r_n^{3/2}}{1-{\cal M}^2/r_n^2} \left(\frac{1}{r_n}-\frac{1}{r_{ix}}\right).
\label{rnOblique}
\end{equation}
This approaches Equation~(\ref{govEq}) only if $r_{ix} \approx r_B$. 

\begin{figure}
\epsscale{0.4}
\plotone{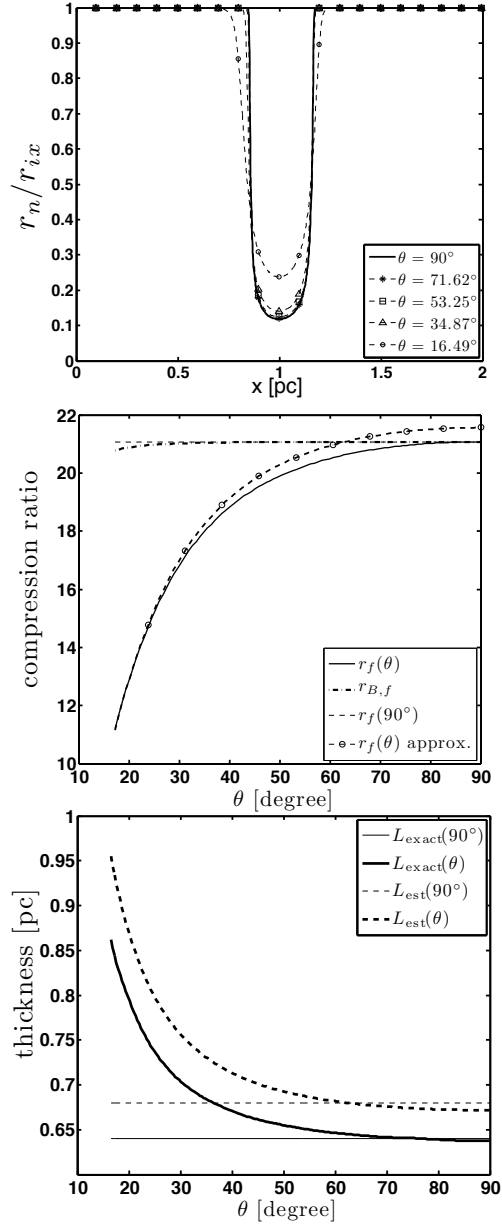}
\caption{The structure, the final compression ratio, and the shock thickness of an oblique C shock with $n_0 = 500~\mathrm{cm^{-3}}$, $B_{y,0} = 5~\mu\mathrm{G}$, $v_0 = 5~\mathrm{km/s}$, and $\chi_{i0} = 5$, as functions of the angle $\theta$ between $\mathbf{B}_0$ and $\mathbf{v}_0$ from $\theta=\theta_\mathrm{min}=16.49^\circ$~to $\theta=90^\circ$. The analytical approximations, Equations~(\ref{rfQuadra}) and (\ref{ObLapprox}), provide good estimates to the exact solutions.\label{ObShock}}
\end{figure}

It is straightforward to show \citep{1991MNRAS.251..119W} that
\begin{equation}
\frac{1}{r_{ix}} = \frac{1}{r_B}\frac{\tan^2\theta + \left(r_n r_B\right)^{-1} + 2\left(r_B-1\right)\left(\beta_y{\cal M}^2\right)^{-1}}{\tan^2\theta + r_B^{-2}};
\label{rviperp}
\end{equation}
evidently $r_{ix} \approx r_B$ for $\theta\rightarrow\pi/2$. By substituting for $r_B$ in terms of $r_n$ using Equation~(\ref{rB}), Equation~(\ref{rviperp}) gives $r_{ix}$ in terms of $r_n$. Using this, Equation~(\ref{rnOblique}) may be integrated. Fig.~\ref{ObShock} shows an example of the C shock structure with varying $\theta$ values. For sufficiently large $\theta$, the shock is changed little with respect to the $\theta=\pi/2$ case. For smaller $\theta$, the structure is quantitatively different, but qualitatively similar. 

The final magnetic compression ratio $r_{B,f}$ is obtained from Equation~(\ref{rviperp}) using $r_f (\theta)\equiv r_{n,f}=r_{ix,f}$ :
\begin{equation}
r_{B,f} = \frac{r_f(\theta)\left(\tan^2\theta - \frac{2}{\beta_y{\cal M}^2}\right)}{\tan^2\theta - \frac{2 r_f(\theta)}{\beta_y{\cal M}^2}}.
\label{rBf}
\end{equation}
From the exact solutions, we know that $r_f (\theta) \leq r_f (\pi/2) \equiv r_{f,90}$, Equation~(\ref{rBf}) therefore suggests that for each model, there is a minimum angle between $\mathbf{B}_0$ and $\mathbf{v}_0$:
\begin{equation}
\theta_\mathrm{min} \approx \tan^{-1}\left(\sqrt{\frac{2 r_{f,90}}{\beta_y{\cal M}^2}}\right) \approx \tan^{-1}\left[\left(\frac{2}{\sqrt{\beta_y}{\cal M}}\right)^{1/2}\right].
\end{equation}
Since, for a given $B_{y,0}$ (or $\beta_y$), small $\theta$ corresponds to large upstream magnetic field strength, a shock is not possible for very small $\theta$. More practically, this can also be written as
\begin{equation}
v_0 \gtrsim \sqrt{2}v_{\mathrm{A,cloud}} \frac{1-\sin^2\theta}{\sin\theta},
\end{equation}
or $\theta > \theta_\mathrm{min}$ (assuming $\sin\theta_\mathrm{min} \ll 1/\sin\theta_\mathrm{min}$) for
\begin{equation}
\sin\theta_\mathrm{min} \sim \sqrt{2} \frac{v_{\mathrm{A,cloud}}}{v_0},
\end{equation}
where $v_{\mathrm{A,cloud}} = B_\mathrm{cloud}/\sqrt{4\pi\rho_0}$. The reason for the condition $\theta > \theta_\mathrm{min}$ is to ensure that the inflow is strong enough to produce shocks in the magnetized gas.

To obtain $r_f (\theta)$, we need to simultaneously solve
\begin{equation}
\frac{{\cal M}^2}{r_f(\theta)} + r_f(\theta) + \frac{r_{B,f}^2}{\beta_y} = {\cal M}^2 + 1 + \frac{1}{\beta_y}
\label{rBf_rf}
\end{equation}
and Equation~(\ref{rBf}), which can only be done numerically. Alternatively, we can also use Equation~(\ref{rBf_rf}) to write (assuming ${\cal M}^2 \gg r_f(\theta) \gg 1$)
\begin{equation}
r_{B,f} \approx \sqrt{\beta_y\left[{\cal M}^2 - r_f(\theta)\right]} \approx \sqrt{\beta_y}{\cal M}\left(1-\frac{r_f(\theta)}{2{\cal M}^2}\right).
\label{rBfApprox}
\end{equation}
Substituting Equation~(\ref{rBfApprox}) into Equation~(\ref{rBf}) gives us a quadratic equation for $r_f(\theta)$:
\begin{equation}
\frac{{r_f}^2(\theta)}{\sqrt{\beta_y}{\cal M}^3 \tan^2\theta} + \left[ \frac{2}{\beta_y{\cal M}^2\tan^2\theta}\left(1-\sqrt{\beta_y}{\cal M}\right)-1-\frac{\sqrt{\beta_y}}{2{\cal M}}\right] r_f(\theta) + \sqrt{\beta_y}{\cal M} = 0.
\label{rfQuadra}
\end{equation}
Since ${\cal M}\gg 1$, keeping only ${\cal M}^{-1}$ terms gives
\begin{equation}
r_f(\theta)\approx \sqrt{\beta_y}{\cal M}\left[\frac{2}{\sqrt{\beta_y}{\cal M}\tan^2\theta}+\frac{\sqrt{\beta_y}}{2{\cal M}}+1\right]^{-1}.
\label{ObrfApprox}
\end{equation}
This is an analytical approximation for $r_f(\theta)$ (see Fig.~\ref{ObShock}). The compression factor $r_{f,90}$ for the case with magnetic field parallel to the shock front ($\tan\theta\rightarrow\infty$) is $r_{f,90} \approx \sqrt{\beta_y}{\cal M}$ (see Equation~(\ref{rfApprox})). Note that for $\tan\theta\gtrsim\left[2/\left(\sqrt{\beta_y}{\cal M}\right)\right]^{1/2}$, $r_f(\theta)\sim r_{f, 90}$. Thus, for all but the smallest angles, oblique shocks have similar compression factor to the $90^\circ$ case with the same $B_{y,0}$.

Since $r_n/r_{ix}$ is small thorough most of the shock just like $r_n/r_B$, we can follow the derivation in Section~\ref{sec:ThickMag} to get the formula for the C shock thickness with different $r_f (\theta)$. Equation~(\ref{rnOblique}) for the shock structure is
\begin{equation}
\frac{d}{dx}\left(r_n+\frac{{\cal M}^2}{r_n}\right) = -D r_n^{1/2}\left(1-\frac{r_n}{r_{ix}}\right),
\end{equation}
similar to Equation~(\ref{drnThickness}). Therefore, as for Equation~(\ref{anaL1}), the oblique C shock thickness can be written as
\begin{equation}
L_\mathrm{est}(\theta) \approx \frac{4{\cal M}^2}{D \left[r_f(\theta)\right]^{1/2}} = \frac{4 v_0}{\alpha\rho_{i,0}\left[r_f\left(\theta\right)\right]^{1/2}}.
\label{ObLapprox}
\end{equation}
An example comparing the approximation Equation~(\ref{ObLapprox}) (using $r_f(\theta)$ from Equation~(\ref{ObrfApprox})) with the exact solution is shown in Fig.~\ref{ObShock}.

Regarding the time-dependent behavior of oblique C shocks, we use convergent flow to produce shocks in numerical simulations. To see how the component of magnetic field parallel to the inflow direction ($B_x$) can affect the evolution of the candidate core material, we fix the values of $n_0$, $v_0$, $B_{y,0}$, $\chi_{i0}$, and choose different values of $\theta$ so that $B_x = B_{y,0}\cot\theta$. Based on our theory, the growth rate of column density $dN(\mathrm{H})/dt$ is proportional to $v_\mathrm{inflow}(\theta) = v_0\left[r_f(\theta) -1\right]/r_f(\theta)$, which should be almost the same for different $\theta$, since $r_f(\theta) \gg 1$. The ambipolar diffusion timescale $t_\mathrm{AD}$ and the final mass-to-flux ratio $\Gamma_\mathrm{final}$, however, should decrease slightly for smaller $\theta$ because of their dependence on $r_f(\theta)$. The generalizations of Equations~(\ref{tADapprox2}) and (\ref{MdBfinal}) are:
\begin{align}
t_\mathrm{AD}(\theta) &\approx \frac{2 \left[r_f(\theta)\right]^{1/2}}{\alpha\rho_{i,0}} = t_{\mathrm{AD}, 90}\left[\frac{r_f(\theta)}{r_{f,90}}\right]^{1/2}, \label{ObtAD}\\
\Gamma_\mathrm{final}(\theta) &\approx 2\pi\sqrt{G}\cdot\frac{2 \rho_0 v_0 \times t_\mathrm{AD}(\theta)}{r_{B,f} B_0} \approx \Gamma_{\mathrm{final}, 90} \left[\frac{r_f(\theta)}{r_{f,90}}\right]^{1/2}\label{ObGamma}
\end{align}
where we apply $r_{B,f}\approx r_{f,90}$ to get the second equation, and Equation~(\ref{GADmodel}) gives $\Gamma_{\mathrm{final}, 90}$. Since $r_f(\theta)/r_{f,90}$ is order-unity unless $\theta$ is extremely small, $\Gamma_\mathrm{final}(\theta)$ is close to $\Gamma_{\mathrm{final}, 90}$ in most cases.

\begin{figure}
\epsscale{0.4}
\plotone{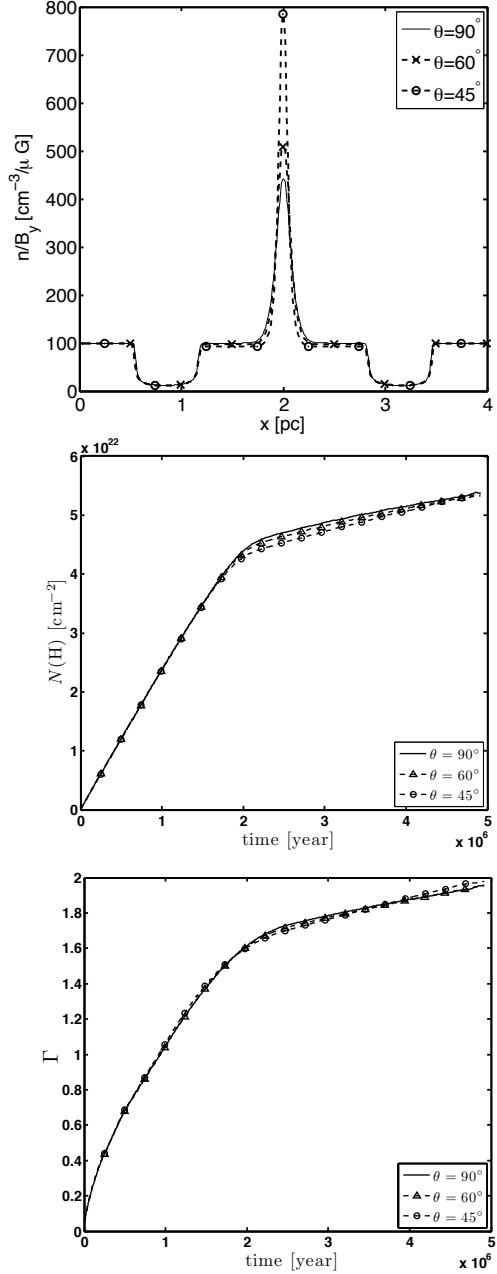}
\caption{The transient behavior and time evolution of the column density and normalized central mass-to-flux ratio in the post-shock gas of oblique C shocks with $n_0 = 500~\mathrm{cm^{-3}}$, $B_{y,0} = 5~\mu\mathrm{G}$, $v_0 = 5~\mathrm{km/s}$, and $\chi_{i0} = 5$. Though the profile of transient central core differs, the growth rate and the final value of $\Gamma$ are very similar in each case. \label{ObCore}}
\end{figure}

The simulation results shown in Fig.~\ref{ObCore} agree with our expectation. The column density grows at an identical rate in all cases (though the shape of the central peak differs from one to another), and the transition happens slightly earlier for smaller $\theta$. There is no obvious difference between the final mass-to-flux ratios in each case, however, since $r_{B,f}$ actually decreases for smaller $\theta$ and makes $\Gamma_\mathrm{final}(\theta)$ slightly larger, thus cancels part of the effect from $r_f(\theta)$. 

In conclusion, these tests show that the evolution of C shock transients to make candidate prestellar cores is not significantly affected by the component of magnetic field parallel to the inflow velocity.

\end{document}